\documentclass[11pt]{article}

\usepackage{amsmath, amssymb, amsfonts, amsthm}
\usepackage[english]{babel}
\usepackage{bm}
\usepackage{mathrsfs}
\usepackage{enumerate}

\numberwithin{equation}{section}

\newtheorem{theorem}{Theorem}[section]
\newtheorem{proposition}[theorem]{Proposition}

\newtheorem{lemma}[theorem]{Lemma}
\newtheorem{definition}[theorem]{Definition}
\newtheorem{remark}[theorem]{Remark}

\renewcommand{\theequation}{\thesection.\arabic{equation}}

\newcommand{\bdm}{\begin{displaymath}}
\newcommand{\edm}{\end{displaymath}}
\newcommand{\bdn}{\begin{eqnarray}}
\newcommand{\edn}{\end{eqnarray}}
\newcommand{\bay}{\begin{array}{c}}
\newcommand{\eay}{\end{array}}
\newcommand{\ben}{\begin{enumerate}}
\newcommand{\een}{\end{enumerate}}
\newcommand{\beq}{\begin{equation}}
\newcommand{\eeq}{\end{equation}}
\newcommand{\bml}[1]{\begin{multline} #1 \end{multline}}
\newcommand{\bmln}[1]{\begin{multline*} #1 \end{multline*}}
\newcommand{\f}{\frac}
\newcommand{\F}{\mathcal{F}}
\newcommand{\pot}{\mathcal{G}_{\la}}

\newcommand{\green}{G_{\la}}
\newcommand{\renform}{\mathcal{F}_{\alpha}^R}

\newcommand{\form}{\F_{\alpha}}
\newcommand{\renqform}{\Phi^{R,\la}_{\al}}

\newcommand{\qform}{\Phi^{\la}_{\alpha}}
\newcommand{\dqform}{\Phi_{\al,\la}^{\mathrm{diag}}}
\newcommand{\oqform}{\Phi_{\la}^{\mathrm{off}}}

\newcommand{\dform}{F^{\mathrm{diag}}}
\newcommand{\oform}{F^{\mathrm{off}}}

\newcommand{\hcm}{H_{\mathrm{cm}}}

\newcommand{\dom}{\mathscr{D}}

\newcommand{\N}{\mathbb{N}}
\newcommand{\R}{\mathbb{R}}

\newcommand{\RT}{\mathbb{R}^3}
\newcommand{\ldf}{L^{2}_{\mathrm{f}}}
\newcommand{\hf}{H^{1}_{\mathrm{f}}}
\newcommand{\hdf}{H^{2}_{\mathrm{f}}}
\newcommand{\hcf}{H^{-1/2}_{\mathrm{f}}}
\newcommand{\hcfpiu}{H^{1/2}_{\mathrm{f}}}

\newcommand{\ci}{\mathbb{C}}

\newcommand{\lf}{\left}
\newcommand{\ri}{\right}
\newcommand{\bra}[1]{\lf\langle #1\ri|}
\newcommand{\ket}[1]{\lf|#1 \ri\rangle}

\newcommand{\xv}{\mathbf{x}}
\newcommand{\Kv}{\mathbf{K}}

\newcommand{\xvi}{\mathbf{x}_i}
\newcommand{\yv}{\mathbf{y}}

\newcommand{\yvi}{\mathbf{y}_i}

\newcommand{\qv}{\mathbf{q}}

\newcommand{\sv}{\mathbf{s}}
\newcommand{\siv}{{\bm \sigma}}
\newcommand{\tv}{\mathbf{t}}
\newcommand{\tav}{{\bm \tau}}

\newcommand{\kv}{\mathbf{k}}
\newcommand{\kkv}{\mathbf{K}}

\newcommand{\pv}{\mathbf{p}}

\newcommand{\zerov}{\mathbf{0}}
\newcommand{\diff}{\mathrm{d}}
\newcommand{\la}{\lambda}

\newcommand{\al}{\alpha}

\newcommand{\OO}{\mathcal{O}}
\newcommand{\dk}{D(\mathbf{K})}

\newcommand{\qga}{Q_{\gamma}}

\newcommand{\qnga}{Q_{n,\gamma}}

\newcommand{\chib}{\Xi_{\beta}}
\newcommand{\chibm}{\Xi_{\beta, l}}

\newcommand{\xit}{\tilde{\xi}}

\newcommand{\half}{\hbox{$\frac{1}{2}$}}

\newcommand{\tx}{\textstyle}
\newcommand{\disp}{\displaystyle}

\newcounter{remark}[section]

\newcommand{\be}{\begin{equation}}
\newcommand{\ee}{\end{equation}}
\newcommand{\bey}{\begin{eqnarray}}
\newcommand{\eey}{\end{eqnarray}}

\newcommand{\bR}{{\mathbb R}}

\newcommand{\sgn}{\mbox{sgn}}

\newcommand{\wt}{\widetilde}

\newcommand{\cP}{{\cal P}}

\newcommand{\supp}{\operatorname{supp}}

\newcommand{\no}{\nonumber}

\input epsf

\newcommand{\donothing}[1]{}
\newcommand{\n}{\noindent}
\newcommand{\vs}{\vspace{0.5 cm}}

\usepackage{bbm}

\begin{document}



\title{Stability for a System of $N$ Fermions Plus a Different Particle with Zero-Range Interactions}

\author{M. Correggi$^{1}$, G. Dell'Antonio$^{2}$, D. Finco$^{3}$, A. Michelangeli$^{4}$, A. Teta$^5$   \\ \\
\small{1. Dipartimento di Matematica, Universit\`a di Roma Tre}\\
\small{Largo San Leonardo Murialdo 1, 00146 Roma, Italy, michele.correggi@gmail.com}\\
\small{2. Dipartimento di Matematica, ''Sapienza'' Universit\`a di Roma}\\
\small{P.le A. Moro 5, 00185 Roma, Italy,  and} \\
\small{Scuola Internazionale di Studi  Superiori Avanzati}\\
\small{Via Bonomea  265, 34136 Trieste, Italy, gianfa@sissa.it}\\
 \small{3. Facolt\`a di Ingegneria, Universit\`a Telematica Internazionale Uninettuno}\\
 \small{Corso V. Emanuele II 39,  00186 Roma, Italy, d.finco@uninettunouniversity.net}\\
 \small{4. Institute of Mathematics, LMU Munich} \\
\small{Theresienstr. 39, 80333 Munich, Germany, michel@math.lmu.de} \\ 
\small{5. Dipartimento di Matematica Pura ed
Applicata, Universit\`a di L'Aquila}\\
\small{Via Vetoio - Loc. Coppito - 67010 L'Aquila, Italy, teta@univaq.it}
\\
}


\maketitle

\begin{abstract}
We study the stability problem for a non-relativistic quantum system in dimension three composed by $N \geq 2$ identical fermions, with unit mass, interacting with  a different particle, with mass $m$, via a zero-range interaction of strength $\alpha \in \bR$. We construct the corresponding  renormalised quadratic (or energy) form  $\form$ and  the so-called Skornyakov-Ter-Martirosyan symmetric extension $H_{\alpha}$, which is  the natural candidate as Hamiltonian of the system. We find a value of the mass $m^*(N)$ such that for $m>m^*(N)$ the form $\form$ is closed and bounded from below. As a consequence, $\form$ defines a unique self-adjoint and bounded from below extension of   $H_{\alpha}$  and therefore the system is stable. 
On the other hand, we also show that the form $\form$ is unbounded from below for $m<m^*(2)$. 
In analogy with the well-known bosonic case, this suggests that 
the system is unstable  for $m<m^*(2)$ and the so-called Thomas effect occurs.
\end{abstract}

\newpage

\vs

\section{Introduction}

The dynamics of a quantum system composed by $N$ particles in $\bR^d$, $d=1,2,3$, interacting via a zero-range, two-body interaction is described by the formal Hamiltonian
\beq\label{hamnd}
\mathcal H =- \sum_{i=1}^N \f{1}{2m_i} \Delta_{\xv_i} + \sum_{\underset{i < j}{i,j=1}}^N \mu_{ij} \,\delta(\xv_i - \xv_j),
\eeq
where $\xv_i \in \bR^d$, $i=1,\ldots,N$,  is the coordinate of the $i$-th particle,   $m_i$ is the corresponding  mass, $\Delta_{\xv_i}$ is the Laplacian relative to  $\xv_i $, and $\mu_{ij} \in \bR$ is the strength  of the interaction between particles $i$ and $j$. To simplify the notation we set $\hbar =1$. 
Formal Hamiltonians of the type \eqref{hamnd} are widely used in physical applications. In particular they are relevant in the study of ultra-cold quantum gases, both in the bosonic and in the fermionic case, in the so-called unitary limit, i.e., for infinite two-body scattering length (see \cite{bh},\cite{wc1},\cite{wc2},\cite{cmp} and references therein).

The first step towards a rigorous approach to the analysis of the model is to give 
the  mathematical definition  of such a Hamiltonian as a self-adjoint operator on the appropriate $L^2$-space. One first notices that  the interaction term in \eqref{hamnd} is effective  only on the hyperplanes $\cup_{i<j} \{\xv_i = \xv_j\}$. This suggests to consider the operator $\dot{\mathcal H}_0$ defined as the free Hamiltonian restricted to a domain of smooth functions vanishing in the neighbourhood of each hyperplane $\{\xv_i=\xv_j\}$. It is easily seen that $\dot{\mathcal H}_0$ is symmetric but not self-adjoint and a trivial self-adjoint extension is the free Hamiltonian on its natural domain. Then, by definition,    any non trivial self-adjoint extension of the operator $\dot{\mathcal H}_0$ is  a  Hamiltonian for a system of $N$ quantum particles in $\bR^d$ with a two-body, zero-range interaction. 
 
The second and more relevant problem  is the concrete construction of such self-adjoint extensions. It turns out that each extension is characterized by a specific generalized boundary condition satisfied by the wave function at the hyperplanes. 
The two most frequently used techniques for the construction are Krein's theory of self-adjoint extensions and the approximation by regularized Hamiltonians, in the sense of the limit of the resolvent or of the quadratic form. The difficulty of the analysis depends on the dimension $d$. For $d=1$ the problem is greatly simplified by the fact that the interaction term is a small perturbation of the free Hamiltonian in the sense of quadratic forms. For $d=2$ a natural class of Hamiltonians with local zero-range interactions was constructed in \cite{DFT} exploiting renormalised quadratic forms and it was also shown that such Hamiltonians are all bounded from below. For $d=3$ the problem is more delicate. In order to illustrate the main point, let us consider the special  case $n=2$  where, in the center of mass  reference frame,  one is reduced to study a one-body problem in the relative coordinate $\xv$  with a  fixed $\delta$-interaction placed at the origin. In this case the problem is completely understood (see, e.g., \cite{al}) and the entire class of  Hamiltonians 
can be explicitly constructed. It turns out that the domain of each Hamiltonian  consists of functions  $ \psi \in L^2(\bR^3) \cap H^2(\bR^3 \setminus \{0\})$ which exhibit the following singular behaviour for $  |\xv| \rightarrow 0$
\beq\label{bc0}
\psi(\xv) = \f{q}{|\xv |} + r  + o(1)\, ,	\hspace{1cm}	\text{with}\;\; \;r=\alpha\, q\, ,
\eeq
where $q \in \ci$ and $\alpha \in \bR$ is a parameter proportional to the inverse of the scattering length. We underline that the relation $r=\alpha q$ in \eqref{bc0} should be understood as the generalized boundary condition satisfied at the origin by all the elements of the domain. In the general case $N>2$ the  characterization of all possible self-adjoint extensions of $\dot{\mathcal H}_0$ is more involved. However,  a  class of extensions  based on the analogy with the case $N=2$ can be explicitly constructed. More precisely, one considers the so-called Skornyakov-Ter-Martirosyan (STM) extension of $\dot{\mathcal H}_0$ which, roughly speaking, is a symmetric operator acting on functions $\psi \in L^2(\bR^{3N}) \cap H^2( \bR^{3N} \setminus \cup_{i<j} \{ \xv_i=\xv_j\})$ 
satisfying the following condition for $|\xv_i - \xv_j| \rightarrow 0$:
\beq\label{bcn}
\psi(\xv_1, \ldots , \xv_n)= \f{q_{ij}}{|\xv_i - \xv_j|} + r_{ij} + o(1)\,,	\hspace{1cm} \text{with} \;\;\; r_{ij}= \alpha_{ij} q_{ij}\, ,
\eeq
where $q_{ij}$ is a  suitable function defined on the hyperplane $\{\xv_i = \xv_j\}$ and $\{\alpha_{ij}\} $ is  a collection of real parameters labelling the extension.  Noticeably, the boundary condition \eqref{bcn} defining the STM extension of $\dot{\mathcal H}_0$ is a  natural  generalization to the case $N>2$ of the condition \eqref{bc0} that characterizes  the two-body case.   Unfortunately, unlike \eqref{bc0}, \eqref{bcn} does not necessarily define a self-adjoint operator. Indeed, for a system of three identical bosons it was shown in \cite{fm} that the STM extension is not self-adjoint and all  its self-adjoint extensions are unbounded from below owing to the presence of an infinite sequence of energy levels $E_k$ going to $-\infty$ for $k \rightarrow \infty$. In \cite{MM} this result was generalized  to the case of three distinguishable  particles with different masses. This kind of instability  is known in the literature as the Thomas effect.  
It should be stressed that the Thomas effect is strongly related to the well-known Efimov effect (see, e.g., \cite{bh}, \cite{adfgl}, \cite{ahkw}) even if, to our knowledge, a rigorous mathematical investigation of this  connection is still lacking. We also mention that if, instead of \eqref{bcn},   one introduces a ``non-local'' boundary condition on the hyperplanes then it is possible to construct a positive Hamiltonian and to study its stability properties for $N$ large (see, e.g., \cite{fs}). In this paper we do not consider this kind of Hamiltonians.
 
It is reasonable to expect that the Thomas effect does not occur if the Hilbert space of states is suitably restricted, e.g., introducing  symmetry constraints on the wave function. A remarkably important constraint is antisymmetry. In fact, a wave function that is antisymmetric under exchange of coordinates of two particles necessarily vanishes at the coincidence points of such two particles, thus making their mutual zero-range interaction ineffective. Analogously, in a mixture of fermions of different species subject to pairwise zero-range interaction, fermions of the same species cannot ``feel'' mutual zero-range interaction and therefore the interaction term in the Hamiltonian is less singular.

In this paper we consider the simplified model consisting of  $N$ identical fermions, with unit mass, and a different particle with mass $m$, interacting with the fermions through a zero-range potential. For such a model only partial results are available and it is remarkable that they strongly depend on the parameters $N$ and $m$. 

Concerning the physical literature, we mention that for $N=2$ it is known (see, e.g., \cite{bh} and references therein)  that for $m < 0.0735= (13.607)^{-1}$ the Thomas effect is present while for $m> 0.0735$ the STM extensions are expected to be bounded from below. More recently (see \cite{cmp}), it was shown by means of analytical and numerical arguments that in the case $N=3$ the Thomas effect occurs for $m< 0.0747= (13.384)^{-1}$. This, in particular, indicates that for $0.0735<m< 0.0747$ there is a sequence of  genuine four-body bound states with energy going to $-\infty$.

From the rigorous point of view, the case $N=2$ was investigated first in \cite{m1} and \cite{MS}, where it was proved that the STM extension is self-adjoint if $m=1$. In \cite{Sh}, the existence of a critical mass  $m^*(2) \simeq 0.0735$ was shown such that for $m<m^*(2)$ any STM extension -- more precisely its restriction to the subspace of angular momentum $l=1$ -- is not self-adjoint and its self-adjoint extensions are unbounded from below. Therefore, the system is unstable and the Thomas effect occurs. This result was extended in \cite{FT}, where  it was shown that the quadratic form associated with the STM extension, restricted to the subspace of angular momentum $l$, is unbounded from below if an explicit condition on $m$ and $l$ is satisfied. Finally, in the case $N\leq 4$ and $m$ sufficiently large it was shown in \cite{m3} that the STM extension is self-adjoint.

In the present work we study the problem for generic $N$ and $m$ and following the line of \cite{DFT} we construct a renormalised quadratic form $\mathcal F_{\alpha}$, $\alpha \in \bR$, which is naturally associated with the STM symmetric extension  $H_{\alpha}$. Here $\alpha$ is the value that all the parameters $\alpha_{ij}$ labelling the STM extension must be equal to as a consequence of the fermionic symmetry. 
Note that in the bosonic case the quadratic form would differ in a sign in front the non-diagonal term (see the remark after \eqref{oqform}).

The first question we address is a non-trivial sufficient condition for the \emph{stability} of the model. As a first main result (Theorem \ref{clbou}) we prove that for any $N$ there is  a value of the mass $m^*(N)>0$  such that $\mathcal F_{\alpha}$ is closed and bounded from below if $m>m^*(N)$. This implies that $\mathcal F_{\alpha}$ is the quadratic form of a \emph{unique} self-adjoint extension of the STM operator $H_{\alpha}$, and this extension is bounded from below. It therefore describes a stable system, where the Thomas effect does not occur.

Such a critical mass  was first conjectured in \cite{m2}  and is precisely the unique root of an explicit equation (see \eqref{Lambda} and \eqref{La=1} in Section \ref{main results: sec}). It turns out that $m^*(N)$ is increasing with $N$ and that the condition $m>m^*(N)$ guarantees the stability also in the limit of infinitely many fermions, provided that the mass of the extra particle scales as $m \propto N$.


The second question we address is a sufficient condition for the \emph{instability} of the model. This can be seen by plugging suitable trial functions into $\mathcal F_{\alpha}$. An attempt in this direction is in \cite[Section 7]{DFT}, but with trial functions that do not satisfy the fermionic symmetry: thus, the result stated there on the unboundedness from below of the quadratic form for $m=1$ and $N$ sufficiently large cannot be considered valid. Our second main result (Theorem \ref{ub1}) fills in this gap and we prove that for any $N\geq 2$ the quadratic form $\mathcal F_{\alpha}$ is unbounded from below for $m<m^*(2)$. In analogy with the bosonic case, we expect that in such a case $H_{\alpha}$ is not self-adjoint and all its self-adjoint extensions are unbounded from below.

Let us make a few remarks on the above-mentioned results. First, we emphasize that in the case $N=2$ we fully characterize stability: the model is stable for $m>m^*(2)$ and unstable $m<m^*(2)$. Whereas the latter was already found in \cite{Sh} by means of the theory of self-adjoint extensions, the former is proved here for the first time.

Second, if $N>2$ we expect the condition $m>m^*(N)$  to be far from optimal for stability. This is due to the crucial role played by the restriction to antisymmetric wave functions (see the discussion in Section \ref{instability: sec}) so that the system might  be stable also if our condition  is violated.  

Third, the fact that the $(N\!+\!1)$-particle system is unstable at least when $m$ is below the \emph{same} threshold $m^*(2)$ for the instability of the $(2\!+\!1)$-particle system has a rather natural interpretation: the instability of a subsystem made of two fermions plus the different particle is responsible for the instability of the whole system. 

We want to mention that in the final stage of the preparation of this work we became aware of a recent paper  \cite{m4} where the case of two fermions plus a different particle is studied using the theory of self-adjoint extensions. We believe that a  comparison with our methods and  results would be of great interest for the further developments of the subject.

The paper is organized as follows. In Section \ref{main results: sec} we introduce the renormalised quadratic form $\mathcal F_{\alpha}$ and the STM extension $H_{\alpha}$ and we formulate our main results. In Section \ref{stability: sec} we give the proof  of Theorem \ref{clbou}. In Section \ref{instability: sec} we give the proof of Theorem \ref{ub1}. In the Appendix we  briefly outline the formal renormalisation  procedure to derive $\form$.

\vs

For the convenience of the reader, we collect here  some useful notation that will be used throughout the paper. We use the notation $ \ldf(\R^{d}) $ (resp. $ \hf(\R^{d}) $, $ \hcf(\R^{d}) $, etc.)  for the space containing totally antisymmetric functions belonging to $ L^2(\R^{d}) $ (resp. $ H^1(\R^{d}) $, $ H^{-1/2}(\R^{d}) $, etc.). We often use the short-hand notation $\|\cdot\|_{L^2}$, $\|\cdot\|_{\ldf}$, etc. for the associated norms $\|\cdot\|_{L^2(\bR^d)}$, $\|\cdot\|_{\ldf(\R^{d})}$, etc.
\newline
For a vector $\xv \in \bR^3$ we set $x = |\xv|$. Moreover we define $ \Kv := \lf( \kv_2,\ldots , \kv_{N-1} \ri) $ and, for $ i = 1, \ldots, N $,
	\bdm
		\breve{\kkv}_i := \lf( \kv_1, \ldots, \kv_{i-1}, \kv_{i+1}, \ldots, \kv_N \ri).
	\edm
For any $f \in L^2(\bR^d)$ the Fourier transform is defined by $ \hat{f}(\kv)=(2 \pi)^{-d/2} \int_{\bR^d} \! \diff \xv \, e^{-i \kv \cdot \xv} f(\xv)\,.$
\newline
The functions $G_{\lambda}$,  $\pot$, $L_{\lambda}$, with $\lambda >0$, and $D(\Kv)$ are defined in \eqref{Green function}, \eqref{pot xi}, \eqref{Lla} and \eqref{Dk} respectively.





\vs

\section{Main results}
\label{main results: sec}

In this section we introduce the quadratic form $\form$,  the STM extension $H_{\alpha}$ and we formulate our main results.

\subsection{The quadratic form $\form$}


\n
For a three dimensional quantum system  composed by $N$ identical fermions, with mass  one, plus a different particles, with mass $m$, with a two-body zero-range interaction    the formal many-body Hamiltonian is 
\beq
	\label{formal Hamiltonian}
	\tilde{H} : = - \frac{1}{2 m}\Delta_{\xv_0} - \f{1}{2} \sum_{i =1}^N \Delta_{\xv_i} + \mu   \sum_{i = 1}^N  \delta(\xv_0 - \xvi)\, ,
\eeq
where $ \xv_i \in \RT $, $ i = 0, \ldots, N $, and $\mu \in \bR$. Introducing the centre of mass and relative coordinates 
\beq
	\lf\{
 		\begin{array}{ll}
			\mathbf{X} : =\displaystyle\frac{1}{m+N} \Big( m\xv_0 + \sum_{i=1}^N \xv_i \Big), 	&	\mbox{}	\\
			\yv_i  : = \xv_0 - \xv_i ,		&	\mbox{for} \:\:\: i = 1,\ldots N\,,
		\eay
	\ri.
\eeq
one obtains
\beq
	\tilde{H} = \hcm + \tx\frac{m+1}{2m}  H 
\eeq
here $ \hcm : =- [2(m+N)]^{-1} \Delta_{\mathbf{X}} $ and
\beq
	\label{cm Hamiltonian}
	H : = - \sum_{i=1}^N \Delta_{\yv_i} -  \frac{2}{m+1} \sum_{i < j} \nabla_{\yv_i}  \cdot \nabla_{\yv_j} +  \mu \sum_{i = 1}^N \delta(\yvi)\,,
\eeq
$\nabla_{\yv_i} $ denoting the gradient with respect to $\yv_i$.  
We also introduce the free Hamiltonian
\beq
	\label{free Hamiltonian}
		\dom(H_0) := H^2_{\mathrm{f}}(\bR^{3N}),	\hspace{1cm}	H_0 : = - \sum_{i=1}^N \Delta_{\yv_i} -  \frac{2}{m+1} \sum_{i < j} \nabla_{\yv_i}  \cdot \nabla_{\yv_j}\,,
\eeq
and its restriction to functions vanishing in a neighbourhood of the hyperplanes 
$ \{\yv_i = 0 \}$
\beq
	\label{free2 Hamiltonian}
\dom(\dot{H}_0)  : = \!\bigg\{ \psi \in \hdf(\R^{3N}) \: \bigg| \! \int_{\R^3} \!\!\diff \kv_i \: \hat\psi(\kv_1, \ldots, \kv_N) = 0 \,, \;\;\;\;i=1,\ldots ,N \bigg\},	\hspace{1cm}	\dot{H}_0 : = H_0\big|_{\dom(\dot{H}_0)}\,.
\eeq

In order to give a rigorous meaning to the formal expression \eqref{cm Hamiltonian} as a self-adjoint operator in  $ \ldf(\R^{3N}) $, one can use Krein's theory to construct self-adjoint extensions of the operator  \eqref{free2 Hamiltonian} (see \cite{m3}). Here, instead, we follow a different approach and we investigate the quadratic form associated with the expectation value $ \bra{\psi} H \ket{\psi} $. However, because of the singularity of the ``potential'' $ \delta(\yv_i) $ the quadratic form has to be defined via a renormalisation procedure in the Fourier space. The idea  of the construction is given in the Appendix (see also  \cite{DFT},\cite{FT}) and here we only give   the final result.

\n
Let us denote for any $\lambda>0$
\beq
	\label{Green function}
	\green(\kv_1, \ldots, \kv_N) : = \bigg[ \sum_{i=1}^N k_i^2 + \frac{2}{m+1}\sum_{i < j} \kv_i \cdot \kv_j + \la \bigg]^{-1}\,,
\eeq
\beq\label{Lla}
L_{\lambda}(\kv_1,\ldots,\kv_{N-1}) := 2 \pi^2 \bigg( \frac{m(m+2)}{(m+1)^2} \sum_{i = 1}^{N-1} k_i^2 + \frac{2m}{(m+1)^2} \sum_{i <j} \kv_i \cdot \kv_j + \la \bigg)^{1/2}\,,
\eeq
and, for any ``charge'' $\xi \in   \hcf(\R^{3N-3})  $,  let us  denote the ``potential'' produced by $\xi$ in the Fourier space by
\beq
	\label{pot xi}
	\lf( \widehat{\pot \xi} \ri)(\kv_1, \ldots, \kv_N) : = \sum_{i=1}^N (-1)^{i+1} \green(\kv_1, \ldots, \kv_N) \, \hat\xi( \breve{\kkv}_i)\,,
\eeq
where  $\breve{\kkv}_i := (\kv_1, \ldots, \kv_{i-1}, \kv_{i+1}, \ldots, \kv_N) $.  Then the quadratic form $\form$  in the Hilbert space $\ldf(\R^{3N})$ is defined as follows
\beq
	\label{form domain}
	\dom(\form) : = \lf\{ \psi \in \ldf(\R^{3N}) \: \Big| \: \exists \xi \in \dom(\qform) \: \;\; \mathrm{s.t.} \;\; \: \phi^{\la} : = \psi - \pot \xi \in \hf(\R^{3N}) \ri\},
\eeq
\beq
	\label{form}
	\form[\psi] : = \F_0[\phi^{\la}] + \la \| \phi^{\la} \|_{L^2(\R^{3N})}^2 - \la \lf\| \psi \ri\|_{L^2(\R^{3N})}^2 + N \qform\lf[\xi\ri],
\eeq
where $\lambda >0$, $ \F_0[\phi] : = \bra{\phi} H_0 \ket{\phi} $ and  $\Phi_{\alpha}^{\lambda}$ is the following form on the charge $ \xi $  
\beq\label{domqform}
 \dom(\qform) := \hcfpiu   (\R^{3N - 3})\,,
\eeq
\beq
	\label{qform}
	\qform[\xi] : = \dqform[\xi] + \oqform[\xi]\,, 
\eeq
\beq	\label{dqform}
	\dqform[\xi] : = \int_{\R^{3N-3}} \diff \kv_1 \cdots \diff \kv_{N-1} \: \big| \hat\xi(\kv_1, \ldots, \kv_{N-1}) \big|^2 \lf[ \al +		
	L_{\lambda}(\kv_1,\ldots,\kv_{N-1}) \ri],
\eeq
\bml{
	\label{oqform}
	\oqform[\xi] : = (N-1) \int_{\R^{3N}} \diff \sv \diff \tv \diff \kv_2 \cdots \diff \kv_{N-1} \:	 \hat\xi^*(\sv, \kv_2, \ldots, \kv_{N-1}) \hat\xi (\tv, \kv_2, \ldots, \kv_{N-1}) \cdot 	\\
	\green(\sv, \tv, \kv_2, \ldots, \kv_{N-1}) .
}

\vspace{0.1cm}

\n
We notice that for $\xi \in  \dom(\qform)$ we have $\pot \xi \in  \ldf(\R^{3N})$ and $\pot \xi \notin \hf(\R^{3N})$. Therefore the decomposition in \eqref{form domain} is meaningful. Moreover, as we shall see in Section \ref{stability: sec}, the form $\qform$ is well-defined on $\dom (\qform)$.

It is worth mentioning that the fermionic constraint implies not only that the wave function is totally antisymmetric but also that the form on the charges \eqref{qform} differs from the bosonic case by a sign in front of the off-diagonal part \eqref{oqform}. This fact results in a weaker effective interaction among the fermions and the stability problem is qualitatively different.

In order to formulate our main results on the form $\form$ we first  introduce our definition of stability parameter. Let us consider the following function
\beq\label{Lambda}
\Lambda (m,N) : =   2\pi^{-1} (N-1)    (m+1)^2 \bigg[ \f{1}{ \sqrt{m(m+2)}} - \arcsin\bigg(\f{1}{m+1}\bigg) \bigg]\,. 
\eeq
It is easy to check that, for $N$ fixed, the function $\Lambda(m,N)$ is decreasing in $m$ and 
\beq
	\lim_{m\rightarrow 0} \Lambda(m,N)=\infty,	\hspace{1cm}	\lim_{m\rightarrow \infty} \Lambda(m,N)=0\,.
\eeq  
Then we have 

\begin{definition}[Stability parameter $m^*(N)$]
	\mbox{}	\\ 
For $N$ fixed, we define $m^*(N)$ as the unique solution to the equation 
\beq\label{La=1}
\Lambda(m,N)=1\,.
\eeq
\end{definition}
\vspace{0.2cm}
\n
If we define $\theta := \arctan \sqrt{1+\f{2}{m}}$,  where $\theta \in (\f{\pi}{4}, \f{\pi}{2})$, a direct computation shows that equation \eqref{La=1} can be equivalently written as
\beq\label{teta}
 \cot 2 \theta + 2 \theta  -\f{\pi}{2}\bigg( 1 - \f{1}{N-1} \cos^2 2 \theta \bigg)=0\,.
\eeq
We remark that   \eqref{teta}  reduces for $N=2$ to the equation found for the critical mass in  \cite[p. 12871, note 10]{pc}. 
We also notice that $m^*(N)$ is positive and increasing with $N$. In particular,  the condition $\Lambda(m,N) <1$ is equivalent to  $m>m^*(N)$. This condition is crucial to guarantee closure and boundedness from below of the form $\form$.

\begin{theorem}[Stability for $ m > m^*(N)$]
\label{clbou}
\mbox{}	\\
Let $N\geq 2$ and  $m>m^*(N)$.  Then  the quadratic form $\form$ is closed and bounded from below. In particular,  it is positive for $\alpha \geq 0$ and 
\beq\label{inff}
\form[\psi] \geq -  \f{\alpha^2}{4 \pi^4 \big(1\!-\!\Lambda(m,N) \big)} \, \|\psi\|^2_{L^2},	\hspace{1cm} \psi \in \dom(\form)\,,
\eeq
for $\alpha <0$.
\end{theorem}

\n
The proof   will be given in Section \ref{stability: sec}. 

Concerning the instability problem, our  result is the following.

\begin{theorem}[Instability for $ m < m^*(2) $]
\label{ub1}
\mbox{}	\\
Let $N\geq 2$ and $m<m^*(2)$. Then the quadratic form $\form$ is unbounded from below for any $\alpha \in \bR$.
\end{theorem}

\n
This theorem generalizes the result obtained for $N=2$ in \cite{FT} and it is proved in Section \ref{instability: sec}. In the case $N=2$, Theorems \ref{clbou} and \ref{ub1} show that the system is stable for $m>m^* (2)$ and unstable for $m<m^*(2)$, therefore our analysis is complete.  When  $N>2$ the problem remains open: no rigorous result is available for $m^*(2)<m<m^*(N)$. 


\n

\vs

\subsection{The STM extension $H_{\alpha}$}

Here we recall the standard definition of the STM extension $H_{\alpha}$ and its connection with the quadratic form $\form$. Moreover, as a direct consequence of Theorem \ref{clbou},  we prove our main result on $H_{\alpha}$. 

As we mentioned in the introduction, $H_{\alpha}$  is    a distinguished  symmetric  extension of the operator  \eqref{free2 Hamiltonian}. For the details of the construction we refer to \cite{m3} and here we only give  the definition
\bml{
	\label{domain Ha 2}
	\dom(H_{\al}) : = \bigg\{ \psi \in \ldf(\R^{3N}) \: \bigg| \: \exists \, \xi \in H^{3/2}_{\mathrm{f}} (\bR^{3N-3}) \: \;\mathrm{s.t.} \;\: \phi_{\la} := \psi - \pot \xi \in \hdf(\R^{3N}),	\\
	\int_{\R^3} \diff \kv_i \: \hat\phi_{\la}(\kv_1, \ldots, \kv_N) =  (\mathcal A^{\lambda}_{\alpha,i}  \hat\xi ) (\breve{\kkv}_i), \, i = 1, \ldots, N \bigg\}\,,
}  
\beq
	\label{action Ha}
	(H_{\al} + \la) \psi = (H_0 + \la) \phi_{\la}\,,
\eeq
where $\lambda>0$ 
 and
\beq	\label{operator A}
(\mathcal A^{\lambda}_{\alpha,i}  \hat\xi ) (\breve{\kkv}_i) 
 : = (-1)^{i+1} \bigg[ \alpha + L_{\lambda}(\breve{\kkv}_i) \bigg] \hat{\xi}(\breve{\kkv}_i) - \!\!  \sum_{j=1, j\neq i}^N \!\!  (-1)^{j+1}  \!\!  \int_{\R^3} \diff \kv_i \:  G_{\lambda} (\kv_1,\ldots, \kv_N) \, \hat{\xi}(\breve{\kkv}_j) \,.
\eeq
The last equality in \eqref{domain Ha 2} should be understood as the boundary condition satisfied by any $\psi \in \dom(H_{\alpha})$. In fact, 
by a straightforward computation, one verifies that such equality implies  the following asymptotic condition for $R\rightarrow \infty$ 
\beq
 \int_{k_i <R} \diff \kv_i \: \hat{ \psi}(\kv_1,\ldots, \kv_N)= 4 \pi R \, (-1)^{i+1} \hat{\xi}(\breve{\kkv}_i) + \alpha \,  (-1)^{i+1} \hat{\xi}(\breve{\kkv}_i) +o(1)
 \eeq
and this is exactly the standard boundary condition, formulated in the Fourier space,  characterizing the STM extension (see \cite{m3}).

The next step is to establish the connection with the form $\form$. For any $\psi \in \dom (H_{\alpha})$ one has
\begin{eqnarray}
&& \bra{\psi} H_{\alpha} +\lambda  \ket{\psi}  = \bra{\psi} H_{0} +\lambda  \ket{\phi_{\lambda}} = \bra{\phi_{\lambda} } H_{0} +\lambda  \ket{\phi_{\lambda}} + 
 \bra{\pot \xi } H_{0} +\lambda  \ket{\phi_{\lambda}}\nonumber\\
 &&=  \bra{\phi_{\lambda} } H_{0} +\lambda  \ket{\phi_{\lambda}} +
 \sum_{i=1}^N (-1)^{i+1} \!\! \int_{\bR^{3N-3}} \!\!\! \diff \breve{\kkv}_i \; \hat{\xi}^*(\breve{\kkv}_i) \int_{\bR^3} \! \diff \kv_i \, \hat{\phi}_{\lambda} (\kv_1,\ldots,\kv_N)\nonumber\\
 &&=\bra{\phi_{\lambda} } H_{0} +\lambda  \ket{\phi_{\lambda}} +
 \sum_{i=1}^N (-1)^{i+1} \!\! \int_{\bR^{3N-3}} \!\!\! \diff \breve{\kkv}_i \; \hat{\xi}^*(\breve{\kkv}_i) (\mathcal A^{\lambda}_{\alpha,i}  \hat\xi ) (\breve{\kkv}_i)\,. 
\end{eqnarray}

\n
Exploiting \eqref{operator A} and the antisymmetry of $\hat{\xi}$  we have
\bml{ \label{afi}
 \sum_{i=1}^N (-1)^{i+1} \!\! \int_{\bR^{3N-3}} \!\!\! \diff \breve{\kkv}_i \; \hat{\xi}^*(\breve{\kkv}_i) (\mathcal A^{\lambda}_{\alpha,i}  \hat\xi ) (\breve{\kkv}_i) \\ 
 = \sum_{i=1}^N \int_{\bR^{3N-3}} \!\!\! \diff \breve{\kkv}_i \; |\hat{\xi}(\breve{\kkv}_i)|^2 \big[ \alpha + L_{\lambda}(\breve{\kkv}_i) \big] 
 -\sum_{i,j=1, i\neq j}^N \!\!\!\! (-1)^{i+j} 
 \int_{\bR^{3N}}\!\! \diff \kv_1\cdots \diff \kv_N \,  \hat\xi^* (\breve{\kkv}_i) \hat\xi (\breve{\kkv}_j) \, G_{\lambda} (\kv_1, \ldots , \kv_N) \\
 = N \dqform [\xi] + N(N-1) \oqform [\xi]\,.
}
Therefore we conclude
\beq\label{hfa}
    \bra{\psi} H_{\alpha} \ket{\psi}    =     \form \lf[\psi \ri], 	\hspace{1cm}	\psi \in \dom(H_{\al})\,.
\eeq

\n
From the above relation and Theorems \ref{clbou}, \ref{ub1},  we obtain the following result.

\begin{theorem}[Self-adjointness and boundedness from below]
\label{thhal}
\mbox{}	\\
One has the  following two possible alternatives:
\ben[(i)]
	\item If $N\geq 2$ and $m>m^*(N)$, then $\form$ defines a unique self-adjoint and bounded from below extension $\hat{H}_{\alpha}$ of the operator $H_{\alpha}$. In particular, $\hat{H}_{\alpha}$ is positive for $\alpha \geq 0$ and 
\beq\label{infsp}
\inf \sigma(\hat{H}_{\alpha}) \geq - \f{\alpha^2}{4 \pi^4 \big(1 - \Lambda(m,N) \big)}, 	\hspace{1cm}	\text{for} \;\;\;\alpha <0.
\eeq 
	\item  If $N \geq 2$ and $m<m^*(2)$, then, for any $\alpha \in \bR$, no self-adjoint extension $ \hat{H}_{\alpha} $ of $ H_{\alpha} $ such that $ \dom(\hat{H}_{\alpha}) \subset  \dom(\form) $ can be both  self-adjoint and bounded from below.
\een

\end{theorem}

\begin{proof}
The assertion (i) immediately follows from Theorem \ref{clbou} and the variational characterization of the infimum of the spectrum of a self-adjoint operator.

The converse (ii) can be obtained by exploiting the Birman-Krein-Vishik theory of positive self-adjoint extensions (see, e.g., \cite{simon}, \cite{P1},\cite{P2}). We omit the details and refer to \cite[Proposition 4.1]{FT} where an analogous results was proven.
\end{proof}

\vs

\n
We remark that, in the case $N\geq 2$ and $m<m^*(2)$, an analogy with the bosonic case suggests that any self-adjoint extension of $H_{\alpha}$ is unbounded from below and the Thomas effect occurs, although such a complete characterization is beyond our purposes. The most natural tool to approach the analysis of this case  is indeed the theory of self-adjoint extensions (see, e.g., \cite{Sh},\cite{m4}).

\n
We conclude this section with a brief comment relative to the special case $N=2$, $m>m^*(2)$. In particular we want to give the explicit characterization of our Hamiltonian $\hat{H}_{\alpha}$ in order to make more transparent a possible comparison of our result with the result in \cite{m4}.

\n
In Section 3 we shall prove the estimate \eqref{cC}, which in particular implies  that the form  $\Phi^{\lambda}_{\alpha}$, defined on $H^{1/2}(\R^3)$, is closed and positive for $\lambda$ sufficiently large. Therefore it defines a positive, self-adjoint operator $\Gamma^{\lambda}_{\alpha}$, $\dom (\Gamma^{\lambda}_{\alpha})$ in $L^2(\R^3)$ explicitly given by
\bdn\label{opGa}
\dom(\Gamma^{\lambda}_{\alpha})= \!\lf\{ \xi \in H^{1/2}(\R^3)\; \Big| \; \Gamma^{\lambda}_{\alpha} \hat\xi \in L^2(\R^3) \ri\},	\nonumber	\\
\lf(\Gamma^{\lambda}_{\alpha} \hat\xi \ri)(\qv)\!= [\alpha + L_{\lambda}(\qv) ] \hat\xi(\qv) + \!\!\int_{\R^3}\!\!\! \diff \pv \, G_{\lambda}(\pv, \qv) \hat\xi(\pv).
\edn
Exploiting this fact and following the same line of \cite[Section5]{DFT}, we obtain
\bml{
	\label{domain Ha3}
	\dom(\hat{H}_{\al})  = \bigg\{ \psi \in \ldf(\R^{6}) \: \bigg| \: \exists \, \xi \in \dom(\Gamma^{\lambda}_{\alpha})  \: \;\mathrm{s.t.} \;\: \phi_{\la} := \psi - \pot \xi \in \hdf(\R^{6}),	\\
	\int_{\R^3} \diff \pv \: \hat\phi_{\la}(\pv, \qv)  =  (\Gamma^{\lambda}_{\alpha}  \hat\xi ) (\qv) \bigg\}\,,
}  
\beq
	\label{action Ha3}
	(\hat{H}_{\al} + \la) \psi = (H_0 + \la) \phi_{\la}.
\eeq
We notice that the above operator differs from the operator \eqref{domain Ha 2}, \eqref{action Ha} only in a larger  class of admissible charges $ \xi $, i.e., the domain $ \dom(\Gamma^{\lambda}_{\alpha}) $ strictly contains $ H^{3/2}(\R^3) $. We also underline that the boundary condition satisfied on the hyperplanes by an element of \eqref{domain Ha3} is the standard STM boundary condition.



\section{Closure and boundedness from below of $ \form $}
\label{stability: sec}

The proof of Theorem \ref{clbou} is based on a careful estimate from below and from above of the form $\qform$ on the charge $\xi$.  
If $ N > 2 $ (with an obvious modification in the case $ N = 2 $) we rewrite 
both the diagonal and the off-diagonal terms  of $\Phi_{\alpha}^{\lambda}$, defined in  \eqref{dqform}, \eqref{oqform},  in a more manageable form (see \cite{m3}), by introducing the change of coordinates
\bdm
	\sv \longrightarrow \siv : = \sv + \frac{1}{m+2} \sum_{i=2}^{N-1} \kv_i, 	\hspace{1cm} 	\tv \longrightarrow \tav : = \tv + \frac{1}{m+2} \sum_{i=2}^{N-1} \kv_i.		
\edm
Then
\beq
	\label{dqform 1}
	\dqform[\xi] = \alpha \, \| \xi\|_{L^2(\R^{3N-3})
	}^2 +	
	2 \pi^2 \int_{\R^{3N-3}} \diff \sigma \diff \Kv \: 
	\big| \xit(\siv, \Kv 
	) \big|^2 \sqrt{\tx\frac{m(m+2)}{(m+1)^2} \sigma^2 + \dk+ \la}\,,
\eeq
\beq
	\label{oqform 1}
	\oqform[\xi] = (N-1) \int_{\R^{3N}} \diff \siv \diff \tav \diff \Kv \: 
	\xit^*(\siv, \Kv 
	) \: \xit (\tav, \Kv 
	) 
	\lf( \sigma^2 + \tau^2 + \tx\frac{2}{m+1} \siv \cdot \tav + \dk + \la \ri)^{-1}\,,
\eeq
where
\beq
\Kv := \kv_2, \ldots , \kv_{N-1}\,,
\eeq
\beq
	\label{xit}
	\xit \lf(\siv, \Kv 
	\ri) := \hat\xi\bigg(\siv - \frac{1}{m+2} \sum_{i=2}^{N-1} \kv_i,  \kv_2, \ldots, \kv_{N-1}
	\bigg)\,,
\eeq
\beq
	\label{Dk}
	\dk : = 
	\frac{m}{(m+1)(m+2)} \bigg( (m+3) \sum_{i=2}^{N-1} k_i^2 + 2 \sum_{i < j} \kv_i \cdot \kv_j \bigg)\,.
\eeq
Notice that $\dk$ satisfies the bound (see \eqref{el12} in the following)
\beq
	\label{estimate Dk}
	\frac{m}{m+1} \sum_{i=2}^{N-1} k_i^2 \leq \dk \leq \frac{m(m+N+1)}{(m+1)(m+2)} \sum_{i=2}^{N-1} k_i^2\,.
\eeq
Last, setting 
\beq
	\siv := \sqrt{\dk + \la} \: \pv,	\hspace{1cm}	\tav : = \sqrt{\dk + \la} \: \qv\,,
\eeq
and
\beq
	\label{charge Qk}
	Q_{\kkv}(\pv) : = (\dk + \la)^{3/4} \: \xit \lf(\sqrt{\dk + \la} \: \pv, \Kv 
	\ri)\,,
\eeq
we obtain
\beq
	\label{qform 1}
	\Phi_0^{\lambda}[\xi]  = \qform[\xi] -  \alpha \| \xi \|_{L^2(\R^{3N-3})}^2 =  \int_{\R^{3N-6}} \diff \Kv
	\sqrt{\dk \!+\! \la} \;\,  F_1 \! \lf[ Q_{\kkv} \ri]\,,
\eeq
where for any $\zeta \geq 0$ we introduced the quadratic form in $L^2(\R^3)$
\beq
	\label{fform}
	\dom(F_{\zeta})\!:=\!\dom(F_1)\!= \! \bigg\{ \! f \! \in \! L^2(\R^3)\,\bigg|\! \int_{\R^3} \!\! \diff \pv \sqrt{p^2 \!+\!1} \, |f(\pv)|^2  <\infty \bigg\},	\quad	F_{\zeta} \lf[f\ri]
	: = \dform_{\zeta} \lf[f\ri] 
	+ \oform_{\zeta}\lf[f\ri],
\eeq
and 
\beq
	\label{dform}
	\dform_{\zeta} \lf[f \ri] 
	= 2 \pi^2 \int_{\R^{3}} \diff \pv  \sqrt{\tx\frac{m(m+2)}{(m+1)^2} p^2 + \zeta} \: \lf| f 
	(\pv) \ri|^2 \,,
\eeq
\beq
	\label{oform}
	\oform_{\zeta} \lf[f \ri] 
	=  (N-1) \int_{\R^{6}} \diff \pv \diff \qv \,	\frac{f 
	^*(\pv) f 
	(\qv)}{  p^2 + q^2 + \tx\frac{2}{m+1} \pv \cdot \qv + \zeta}\,   .
\eeq

\vspace{0.3cm}
\n
Using the representation \eqref{qform 1}, \eqref{fform}, \eqref{dform}, \eqref{oform} we obtain the following estimate for $\Phi_0^{\lambda}$, which is the crucial ingredient for the proof of Theorem \ref{clbou}.

\begin{proposition}[Upper and lower bounds for $ \Phi_0^{\lambda} $]
\label{stiqform}
\mbox{}	\\
For any $\xi \in \dom(\qform)$ we have
\beq\label{abqform}
\Big(1-\Lambda(m,N) \Big)\; \Phi_{0,\lambda}^{\mathrm{diag}}  [\xi] \;  \leq \;  \Phi_0^{\lambda} [\xi]  \; \leq \; \Big( 1 + \Gamma(m,N)\Big) \;\Phi_{0,\lambda}^{\mathrm{diag}} [\xi]\,, 
\eeq
where
\beq\label{Gamma}
\Gamma(m,N) : = \f{(N\!-\!1) (m\!+\!1)^2}{\sqrt{m(m\!+\!2)}} \arcsin \lf( \f{1}{m\!+\!1} \ri)\,.
\eeq
\end{proposition}

\vspace{0.3cm}

The proof  is based on a careful analysis of the form $F_{\zeta}$, $\zeta \in [0,1]$,  reduced to   each subspace with fixed  angular momentum $l$ and it is postponed to the end of this section. First we introduce some useful  notation and we prove some preliminary lemmas.

For any $f\in L^2(\R^3)$ we consider the  expansion
\beq\label{exsh}
f(\pv) = \sum_{l=0}^{\infty}\sum_{m=-l}^l f_{lm}(p) Y_{l}^m(\theta_p, \phi_p)\,,
\eeq
where $\pv=(p, \theta_p,\phi_p)$ in spherical coordinates and $Y_{l}^m$ denotes the spherical harmonics of order $l,m$. We notice that $f\in \dom(F_1)$ is equivalent to
\beq
\sum_{l=0}^{\infty}\sum_{m=-l}^l \int_0^{\infty}\!\!dp\, p^2 \sqrt{p^2 +1}\,  \lf|f_{lm}(p)\ri|^2 <\infty\,.
\eeq
Moreover, we denote by $P_l$ the Legendre polynomial of order $l=0,1,\ldots$ explicitly given by
\beq\label{leg}
P_l(y) = \f{1}{2^l l!} \f{d^l}{dy^l} (y^2 -1)^l \,,	\hspace{1cm}	 y \in [-1,1]\,.
\eeq
 In the first lemma we decompose $F_{\zeta}$ in each subspace of fixed angular momentum $l$.

\begin{lemma}[Decomposition of $ F_{\zeta} $]
\label{lemma0}
\mbox{}	\\
For $f\in \dom(F_{1})$ we have
\beq	\label{Fzl}
F_{\zeta}\lf[f \ri] \; =\;  	\sum_{l=0}^{\infty}\sum_{m=-l}^l  G_{\zeta,l} \lf[ f_{lm} \ri] \;  =:  \; \sum_{l=0}^{\infty}\sum_{m=-l}^l  \lf( G^{\mathrm{diag}}_{\zeta} \lf[f_{lm} \ri] + G^{\mathrm{off}}_{\zeta,l} \lf[f_{lm} \ri]  \ri)\,,
\eeq  
where for $g \in L^2((0,\infty), \, p^2 \sqrt{p^2+1}\,dp)$
\begin{eqnarray}
	\label{G diag off}
&&G^{\mathrm{diag}}_{\zeta} \lf[ g \ri] : = 2 \pi^2 \!\! \int_0^{\infty} \!\!\! dp \, p^2 \sqrt{\frac{m(m+2)}{(m+1)^2} p^2 +\zeta}\, \, |g(p)|^2\,, \\
&& G^{\mathrm{off}}_{\zeta,l} \lf[ g\ri] : = 2\pi (N-1) \!\int_0^{\infty} \!\!\!\! \!dp\!\!\int_0^{\infty}\!\!\!\!\! dq\, p^2 g^*(p) \, q^2 g(q) \!\int_{-1}^1 \!\!\!\! dy\, \frac{P_l(y)}{p^2 + q^2 +\frac{2}{m+1} p q y +\zeta}\,.
\end{eqnarray}
\end{lemma}

\begin{proof}
For a given $f\in \dom(F_{1})$ we consider the expansion \eqref{exsh}. From \eqref{dform} we see that  $\dform_{\zeta}[f]= \sum_{l=0}^{\infty} \sum_{m=-l}^l G^{\text{diag}}_{\zeta}[f_{lm}]$.  Concerning the off-diagonal term \eqref{oform}, 
we denote by $\theta_{pq}$ the angle between the vectors $\pv$ and $\qv$ and we consider the following expansion in Legendre polynomials:
\begin{eqnarray}\label{exle}
&&\f{1}{p^2 +q^2 + \f{2}{m+1} pq \cos \theta_{pq} +\zeta} = \sum_{l=0}^{\infty}\f{2l+1}{2} \int_{-1}^1 \!\!\!\! dy\, \frac{P_l(y)}{p^2 + q^2 +\frac{2}{m+1} p q y +\zeta}  \,\, P_l(\cos \theta_{pq})\nonumber\\
&&= \sum_{l=0}^{\infty} 2\pi  \int_{-1}^1 \!\!\!\! dy\, \frac{P_l(y)}{p^2 + q^2 +\frac{2}{m+1} p q y +\zeta}  \sum_{m=-l}^{l} Y_{l}^{m *} (\theta_p,\phi_p) Y_{l}^m(\theta_q,\phi_q)\,.
\end{eqnarray}
In the last line we used the addition formula for spherical harmonics (see, e.g., \cite[Eq. (8.814)]{GR}). From (\ref{exle}) we obtain $\oform_{\zeta}[f]=   \sum_{l=0}^{\infty} \sum_{m=-l}^l G^{\text{off}}_{\zeta,l}[f_{lm}]$.
\end{proof}

In the next lemma we give a new representation of $G^{\text{off}}_{\zeta,l}$ which is particularly useful to  control $G^{\text{off}}_{\zeta,l}$ in terms of $G^{\text{off}}_{0,l}$ for any $\zeta>0$. 

\begin{lemma}[Estimates for $ G^{\mathrm{off}}_{\zeta,l}$]
\label{lemma01}
\mbox{}	\\
The form $G^{\mathrm{off}}_{\zeta,l}$ can be written as
\begin{eqnarray}\label{rapof}
&&G^{\mathrm{off}}_{\zeta,l} \lf[ g\ri] = \sum_{k=0}^{\infty} B_{l,k} \int_0^{\infty}\!\!\! d\nu \, \nu^k e^{-\zeta \nu} \lf| \int_0^{\infty}\!\!\! dp \, g(p) \,p^{2+k} e^{-\nu p^2} \ri|^{2}\,,
\end{eqnarray}
where
\begin{eqnarray}\label{Blk}
&&B_{l,k}=
		\begin{cases}
			 \f{2\pi (N-1)}{2^l l! \,k!}  \lf(\! \frac{-2}{m\!+ \!1} \!\ri)^{\!k}  \disp\int_{-1}^1\!\!\! dy \, (1-y^2)^l \frac{d^l}{dy^l} y^k 	& \mbox{if} \:\:\:\: l \leq k\,,	\\
			0	&	\mbox{otherwise}\,.
		\end{cases}
\end{eqnarray}

\n 
Moreover for any $\zeta > 0$ we have
\begin{eqnarray}\label{stle}
&&0\leq  G^{\mathrm{off}}_{\zeta,l} \lf[ g\ri] \leq G^{\mathrm{off}}_{0,l} \lf[ g\ri] \hspace{1cm} \text{for $l$ even}\,,\\
&&\nonumber\\
&&G^{\mathrm{off}}_{0,l} \lf[ g\ri] \leq G^{\mathrm{off}}_{\zeta,l} \lf[ g\ri] \leq 0 \hspace{1cm} \text{for $l$ odd}\,.\label{stlo}
\end{eqnarray}
\end{lemma}

\begin{proof}
Using the expansion
\beq
\f{1}{p^2+q^2+ \f{2}{m+1} pqy + \zeta}= 
 \f{1}{p^2+q^2+\zeta}  
\sum_{k=0}^{\infty} \lf(\!  \f{-2}{m+\!1}\, \f{pqy}{p^2+q^2 +\zeta} \!\ri)^{\!\!k}
\eeq
and formula (\ref{leg}), we obtain
\beq\label{exse1}
G^{\mathrm{off}}_{\zeta,l}\lf[ g \ri]= \f{2\pi (N-1)}{ 2^l l! } \!\sum_{k=0}^{\infty}  \lf(\! \f{-2}{m+1} \!\ri)^{\!\!k} \!\! \int_0^{\infty}\!\!\! \!dp \!\int_0^{\infty}\!\!\!\! dq\, \f{p^{2+k} g^*(p) q^{2+k} g(q)  }{(p^2+q^2+\zeta)^{k+1}}\!
\int_{-1}^1\!\!\!\!dy\, y^k \f{d^l}{dy^l} (y^2-1)^l \,.
\eeq
Integrating  by parts $l$ times we find
\begin{eqnarray}\label{exse2}
&&G^{\mathrm{off}}_{\zeta,l}\lf[ g \ri]=\sum_{k=0}^{\infty} B_{l,k} \, k! \!\int_0^{\infty}\!\!\! \!dp \!\int_0^{\infty}\!\!\!\! dq\, \f{p^{2+k} g^*(p) q^{2+k} g(q)  }{(p^2+q^2+\zeta)^{k+1}}\,.
\end{eqnarray}
Finally we use the identity
\beq\label{iden}
\f{k!}{(p^2+q^2+\zeta)^{k+1}} = \int_0^{\infty}\!\!\!\! d\nu \, \nu^{k} \,e^{-(p^2+q^2+\zeta)\nu}
\eeq
in (\ref{exse2}) and we obtain (\ref{rapof}). Let us fix $l$ even. Then the integral in (\ref{Blk}) is different from zero only if $k$ is even and this implies that $G^{\mathrm{off}}_{\zeta,l}$ is positive and the estimate (\ref{stle}) holds. Analogously, when $l$ is  odd  the integral in (\ref{Blk}) is different from zero only if $k$ is odd and therefore $G^{\mathrm{off}}_{\zeta,l}$ is negative and (\ref{stlo}) holds.
\end{proof}


\noindent
Now we study the form $G_{0,l} = G^{\text{diag}}_{0} + G^{\text{off}}_{0,l}$ and we show that it can be diagonalized for each $l$. 
\begin{lemma}[Diagonalization of $G_{0,l}$]
\label{lemma1}
	\mbox{}	\\
	 For any $g \in L^2((0,\infty), p^2\sqrt{p^2+1}\, dp)$ we have
\begin{eqnarray} \label{formula0}
&&G^{\mathrm{diag}}_0 [g] =   2 \pi^2 \f{\sqrt{m(m+2)}}{m+1}  \int_{\R} \diff k \,  \big|  g^{\sharp} (k) \big|^2\,, \\
&&G^{\mathrm{off}}_{0,l}[g] = \int_{\R} \diff k \, S_l (k) \big|  g^{\sharp} (k) \big|^2\,,
\label{formula1} 
\end{eqnarray}
where
\beq \label{formula2}
 g^{\sharp} (k) : = \frac{1}{\sqrt{2\pi}} \int_{\R} \diff x \, e^{-ikx}\, e^{2x} \, g(e^x)
\eeq
and 
\beq \label{formula3}
S_l (k) = 2\pi^2 (N-1) \int_{-1}^1 \diff y \, P_l (y) 
\frac{\sinh \lf( k \arccos \frac{y}{m+1} \ri) }{ \sin  \lf(  \arccos \frac{y}{m+1} \ri) \, \sinh(\pi k)}\,.
\eeq
\end{lemma}
\begin{proof}
The proof of (\ref{formula0}) is straightforward. To prove (\ref{formula1}) we first introduce the new integration variables
 $p=e^{x_1}$ and $q=e^{x_2}$, so that  the form reads
\begin{equation}\label{for2}
\begin{split}
G^{\mathrm{off}}_{0,l} \lf[g \ri]\;&=\;2 \pi(N-1) 
\int_{\R}  dx_1 dx_2\,\,e^{3 x_1}{g}^* (e^{x_1})e^{3 x_2} g(e^{x_2})
\int_{-1}^1 dy  \f{  P_l(y)   }{ e^{2 x_1} + e^{2x_2} +\f{2y}{m+1} e^{x_1 +x_2}  } \\
& = \; \pi(N-1) 
\int_{\R}  dx_1 dx_2\,\,e^{2 x_1}{g}^* (e^{x_1})e^{2 x_2} g(e^{x_2})
\int_{-1}^1 dy  \f{  P_l(y)   }{ \cosh (x_1-x_2) +\f{y}{m+1}  }\,.
\end{split}
\end{equation}
The kernel in (\ref{for2}) is a convolution kernel and therefore it can be diagonalized by means of the Fourier transform. Using the explicit Fourier transform of the kernel (see, e.g., \cite{erdely}) we finally arrive at  \eqref{formula1}.
\end{proof}

Owing to the previous lemma, the problem of finding  bounds for  the form $G^{\text{off}}_{0,l} $ is reduced to finding 
bounds for the function $S_l(k)$. 
Taking into account the identity $\arccos z = \f{\pi}{2} - \arcsin z$ and the parity of $P_l$,  we represent $S_l(k)$ as
\begin{equation}\label{reps}
S_l(k) =
\begin{cases}
   \displaystyle  - \pi^2 (N-1) 
\int_{-1}^1 dy \, P_l(y) \f{\sinh \lf( k \arcsin  \f{ y }{m+1}\ri) }{ \cos \lf(  \arcsin  \f{ y }{m+1}\ri)\sinh \lf(\f{\pi}{2} k\ri) \, } & \text{for} \;  l \text{ odd}, \\
  \displaystyle  \pi^2(N-1) 
\int_{-1}^1 dy  \, P_l(y) \f{\cosh \lf( k \arcsin  \f{ y }{m+1}\ri) }{ \cos \lf(  \arcsin  \f{ y }{m+1}\ri)\cosh \lf(\f{\pi}{2} k\ri) \, } & \text{for} \;  l \text{ even}\,.
\end{cases}
\end{equation}
It is evident from this representation that $S_l (k)$ is for any $l \geq 0 $ an even $C^{\infty}$-function of $k$ with $\lim_{k \rightarrow \infty} S_l (k)=0$.   
Before discussing upper and lower bound of $S_l (k) $, we shall prove the following elementary lemma.
\begin{lemma}[Taylor expansions of $ \wt S^{\mathrm{o}}_k  $ and $ \wt S^{\mathrm{e}}_k $]
\label{forever}
	\mbox{}	\\
For any fixed $k\geq 0$ the following functions 
\beq\label{s0y}
\wt S^{\mathrm{o}}_k (y) = \f{\sinh \lf( k \arcsin  \f{ y }{m+1}\ri) }{ \cos \lf(  \arcsin  \f{ y }{m+1}\ri) \, } \,,	\hspace{1cm}	 \wt S^{\mathrm{e}}_k (y) = \f{\cosh \lf( k \arcsin  \f{ y }{m+1}\ri) }{ \cos \lf(  \arcsin  \f{ y }{m+1}\ri) \, }
\eeq
have a Taylor expansion in the variable $y \in [-1,1]$ with positive coefficients.
\end{lemma}
\begin{proof}
Let $\cP$ be the set of functions whose Taylor expansion has positive coefficients.
First note that $\arcsin y$, $\sinh y$ and $\cosh y$ belong to $\cP$. 
The derivative is a linear automorphism of $\cP$ and therefore 
\[
\frac{d}{dy} \arcsin y = \frac{1}{\sqrt{1-y^2} } = \frac{1}{\cos \arcsin y} \in \cP\,.
\]
Moreover, $\cP$ is invariant under dilations, multiplications and compositions of functions in $\cP$.
Thus, $\wt S^{\mathrm{o}}_k$ and  $\wt S^{\mathrm{e}}_k $ belong to $ \cP$.
\end{proof}

\n
In the next lemma we compute  lower and upper bounds for $S_l (k) $.
\begin{lemma}[Bounds for $S_l (k) $]
 \label{lem4}
	\mbox{}	\\
For  any $k \in \R$ we have
\begin{eqnarray}\label{boe}
0  \; \leq  \;           S_l(k)  \;  \leq     \;   2 \pi^2 (N\!-\!1)(m\!+\!1) \arcsin \lf(\f{1}{m\!+\!1}\ri) &	\hspace{0,7cm}	  & \text{for $l$ even}	\\
\label{boo}
-4\pi (N\!-\!1) (m\!+\!1) \bigg[ 1- \sqrt{m(m\!+\!2)} \, \arcsin  \lf(\f{1}{m\!+\!1}\ri) \bigg]
\; \leq S_l(k)  \;  \leq 0 &	\hspace{0,7cm} & \text{for $l$ odd}\,.
\end{eqnarray}
\end{lemma}
\begin{proof}
Let us prove \eqref{boo} first. The upper bound follows from \eqref{stlo} and \eqref{formula1}. 
As for the lower bound,  by means of \eqref{s0y}  we write
\beq \label{tonno}
S_l (k) = - \frac{\pi^2 (N-1)}{ \sinh \lf(\f{\pi}{2} k\ri)    }\int_{-1}^{+1} \diff y \, P_l (y) \, \wt S^{\mathrm{o}}_k (y) \,.
\eeq
The first step is to  prove that $S_l (k)$ is an increasing function of $l$ for any  fixed $k$. 
From \eqref{tonno},  using \eqref{leg} and  integrating by parts, we obtain
\begin{align} \label{tonno2}
S_{l+2}(k) &= - \frac{\pi^2 (N-1)}{ \sinh \lf(\f{\pi}{2} k\ri)    } \frac{1}{2^{l+2} (l+2)!}
\int_{-1}^{+1} \diff y \, \frac{d^{l+2}}{dy^{l+2} }(y^2-1)^{l+2}  \wt S^{\mathrm{o}}_k (y) \no \\
& = \frac{\pi^2 (N-1)}{ \sinh \lf(\f{\pi}{2} k\ri)    } \frac{1}{2^{l+2} (l+2)!}
\int_{-1}^{+1} \diff y \, \frac{d^{2}}{dy^{2} }(y^2-1)^{l+2} \frac{d^{l} \wt S^{\mathrm{o}}_k}{dy^{l} }(y) \no \\
& = \frac{\pi^2 (N-1)}{ \sinh \lf(\f{\pi}{2} k\ri)    } \frac{1}{2^{l+2} (l+2)!}
\int_{-1}^{+1} \diff y \, \lf[(l+2)(l+1)(y^2-1)^{l}4y^2 + 2(l+2)(y^2-1)^{l+1}    \ri] 
\frac{d^{l} \wt S^{\mathrm{o}}_k}{dy^{l} }  (y) \no \\
&=  \frac{\pi^2 (N\!-\!1)}{ \sinh \lf(\f{\pi}{2} k\ri)    } \frac{1}{2^{l} l!}   
\int_{-1}^{+1} \!\!\!\diff y \, (y^2-1)^{l} 
\frac{d^{l} \wt S_k}{dy^{l} }  (y) + \frac{\pi^2 (N\!-\!1)}{ \sinh \lf(\f{\pi}{2} k\ri)    } \f{1\!+2(l\!+\!1)}{2^{l\!+\!1} (l\!+\!1)!} \int_{-1}^{+1} \!\!\!\diff y \, (y^2-1)^{l+1} 
\frac{d^{l} \wt S^{\mathrm{o}}_k}{dy^{l} }  (y)
\no\\
&= S_l(k)  + \frac{\pi^2 (N\!-\!1)}{ \sinh \lf(\f{\pi}{2} k\ri)    } \f{1\!+2(l\!+\!1)}{2^{l\!+\!1} (l\!+\!1)!} \int_{-1}^{+1} \!\!\!\diff y \, (y^2-1)^{l+1} 
\frac{d^{l} \wt S^{\mathrm{o}}_k}{dy^{l} }  (y) \,. 
\end{align}
By Lemma \ref{forever}, and taking into account that $l+1$ is even, we deduce that the last integral in \eqref{tonno2} is positive and therefore  we conclude  $S_{l+2} (k) \geq S_l(k)$. 
This means that it is sufficient to find the minimum of $S_1 (k)$. 
From  \eqref{reps} we have 
\beq \label{shotgun}
 S_1 (k) = 
-2\pi^2 (N-1) 
\int_{0}^1 \diff y \, \f{y}{\cos \lf(  \arcsin  \f{ y }{m+1}\ri)}   \f{\sinh \lf( k \arcsin  \f{ y }{m+1}\ri) }{\sinh \lf(\f{\pi}{2} k\ri) \, }\,.
\eeq
We know that $S_1(0)<0$ and $\lim_{k \rightarrow \infty} S_1(k)=0$. Moreover,  the derivative of $S_1(k)$ does not vanish for $k>0$, which follows from the fact that the derivative of the function
\[
\f{\sinh ak}{\sinh bk}\,,	\hspace{1cm} 0<a<b,
\]
does not vanish for $k>0$. Therefore, $S_1(k)$ is monotone increasing when $k>0$ and attains    its minimum at $k=0$. Thus,
\begin{equation}\label{S10}
\begin{split}
S_l (k) \;\geq\; S_1 (k) \;\geq\; S_1 (0)\;&=\;-4\pi (N-1) \int_{0}^1 \diff y \, y
\f{ \arcsin  \f{ y }{m+1} }{ \cos \lf(  \arcsin  \f{ y }{m+1}\ri) } \\
&=\;-4\pi (N-1) (m+1)^2 \int_0^{\arcsin \f{ 1 }{m+1} } \diff z \, z \sin z  \\
&=\;-4\pi (N-1) (m+1) \bigg[1- \sqrt{m(m+2)} \, \arcsin \lf(\f{1}{m+1}\ri) \bigg]
\end{split}
\end{equation}
and \eqref{boo} is proved. The proof of \eqref{boe}  is completely analogous. In this case the lower bound follows from \eqref{stle} and \eqref{formula1}.  Using the representation
\beq
S_l (k) =  \frac{\pi^2 (N-1)}{ \sinh \lf(\f{\pi}{2} k\ri)    }\int_{-1}^{1} \diff y \, P_l (y) \, \wt S^e_k (y)
\eeq
obtained from \eqref{s0y}, one sees that $S_{l+2}(k) \leqslant S_l(k)$ and therefore  it is enough to consider $S_0(k)$. Since
\beq
S_0(0)= \pi^2 (N\!-\!1) \!\int_{-1}^{1}\!\!\! \diff y \f{1}{\cos \left(\! \arcsin \f{y}{m+1} \! \right)} = 2 \pi^2 (N\!-\!1)(m\!+\!1) \arcsin \lf( \f{1}{m\!+\!1} \ri) >0\,,
\eeq
$\lim_{k \rightarrow \infty} S_0(k)=0$, and the derivative of $S_0(k)$ does not vanish for $k>0$, we deduce that $S_0(0)$ is the maximum of $S_0(k)$.
\end{proof}

Using the results of the previous lemmas we can finally 
prove Proposition \ref{stiqform}. 
 
\begin{proof}[Proof of Proposition \ref{stiqform}]  We prove first the estimate from below in \eqref{abqform}. From Lemmas \ref{lemma0},  \ref{lemma01}, \ref{lemma1}, \ref{lem4} we have
\begin{equation}\label{sotto}
\begin{split}
\oform_1[f]\;&=\;\sum_{\substack{l,m \\ l\textrm{ even}}}
 G^{\mathrm{off}}_{1,l} \lf[f_{lm} \ri]   + \sum_{\substack{l,m \\ l\textrm{ odd}}}
 G^{\mathrm{off}}_{1,l} \lf[f_{lm} \ri] \;\geq\;\sum_{\substack{l,m \\ l\textrm{ odd}}}
 G^{\mathrm{off}}_{0,l} \lf[f_{lm} \ri] \;=\;\sum_{\substack{l,m \\ l\textrm{ odd}}} \int_{\R} dk\, S_l(k) \, \big| f^{\sharp}_{lm} (k) \big|^2 \\
&\geq\; -4\pi (N\!-\!1) (m\!+\!1)  \bigg[ 1 - \sqrt{m(m+2)} \, \arcsin \lf( \f{1}{m+1} \ri) \bigg]
  \;  \sum_{l,m} \int_{\R} dk\,  \big| f^{\sharp}_{lm} (k) \big|^2 \\
&=\;- \Lambda(m,\!N) \, \f{2 \pi^2 \sqrt{m(m+2)}}{m+1}  \sum_{l,m} \int_{\R} dk\,  \big| f^{\sharp}_{lm} (k) \big|^2
\end{split}
\end{equation}
where $\Lambda(m,\!N)$ is defined in \eqref{Lambda}. Then, by \eqref{formula0},
\beq\label{sotto1}
\oform_1[f] \geq -\Lambda(m,\!N) \! \sum_{l,m} G^{\mathrm{diag}}_{0} [f_{lm}]  \geq -\Lambda(m,\!N) \! \sum_{l,m} G^{\mathrm{diag}}_{1} [f_{lm} ] =  -\Lambda(m,\!N) \, \dform_1[f]\,.
\eeq
From the representation \eqref{qform 1} and from \eqref{sotto1} we have
\beq
\Phi_0^{\lambda} [\xi] \geq \big(1-\Lambda(m,N)\big) \int_{\bR^{3N-6}} \!\!\! \diff \Kv \, \sqrt{D(\Kv) +\lambda} \, F_1^{\mathrm{diag}} [Q_{\Kv}] = \big(1-\Lambda(m,N)\big)\, \Phi_{0,\lambda}^{\mathrm{diag}} [\xi]
\eeq
and the lower bound is proved. An analogous proof yields the upper bound in \eqref{abqform}. We have
\begin{equation}\label{sopra}
\begin{split}
\oform_1[f]\;&\leq\;2 \pi^2 (N\!-\!1)(m\!+\!1) \arcsin \lf( \f{1}{m\!+\!1} \ri) \sum_{l,m} \int\!\! dk\,  | f^{\sharp}_{lm} (k)|^2 \\
&=\;\f{(N\!-\!1) (m\!+\!1)^2}{\sqrt{m(m\!+\!2)}} \arcsin \lf(\f{1}{m\!+\!1} \ri) \sum_{l,m} G^{\mathrm{diag}}_{0} [f_{lm}] \\
&\leq\;\f{(N\!-\!1) (m\!+\!1)^2}{\sqrt{m(m\!+\!2)}} \arcsin \lf( \f{1}{m\!+\!1} \ri) \sum_{l,m} G^{\mathrm{diag}}_{1} [f_{lm}] \\
&=\;\f{(N\!-\!1) (m\!+\!1)^2}{\sqrt{m(m\!+\!2)}} \arcsin \lf( \f{1}{m\!+\!1} \ri) \dform_1[f]
\end{split}
\end{equation}
which, together with \eqref{qform 1}, yields the upper bound for $\Phi_0^{\lambda}$. 
\end{proof}

Let us briefly comment on the result of Proposition \ref{stiqform}. By means of the elementary estimate
\beq\label{el12}
-\f{1}{2} \sum_{i=1}^{N-1} k_i^2 \leq \sum_{i<j} \kv_i \cdot \kv_j \leq \f{N-2}{2} \sum_{i=1}^{N-1} k_i^2\,,
\eeq
we find
\bml{\label{fiup}
	\qform [\xi] = \alpha \|\xi\|^2_{L^2} +\Phi_0^{\lambda} [\xi]  \leq \alpha \|\xi\|_{L^2} + \Big( 1 + \Gamma(m,N)\Big) \;\Phi_{0,\lambda}^{\mathrm{diag}} [\xi]	\\
	\leq \int_{\bR^{3N-3}} \!\! \! \diff \kv_1 \cdots \diff \kv_{N-1} \! \big| \hat\xi(\kv_1, \ldots, \kv_{N-1}) \big|^2 \! \bigg[ \al \!+\! 2 \pi^2  \Big(\! 1\! +\! \Gamma(m,N)\Big)  \bigg( \frac{m(m\!+\!N)}{(m+1)^2} \sum_{i = 1}^{N-1} k_i^2 + \la \bigg)^{\!\!1/2}  \bigg]
}
and
\beq\label{fido}
 \qform [\xi] \geq \int_{\bR^{3N-3}} \!\! \! \diff \kv_1 \cdots \diff \kv_{N-1} \! \big| \hat\xi(\kv_1, \ldots, \kv_{N-1}) \big|^2 \! \bigg[ \al + 2 \pi^2  \Big(\! 1 - \Lambda(m,N)\Big)  \bigg( \frac{m}{m\!+\!1} \sum_{i = 1}^{N-1} k_i^2 + \la \bigg)^{\!\!1/2}  \bigg]\,.
\eeq 
From \eqref{fiup}, \eqref{fido}, choosing $\lambda$ sufficiently large if $\alpha <0$, we conclude   
\beq\label{cC}
c\, \Big( 1- \Lambda(m,N) \Big)  \|\xi \|_{\hcfpiu} \leq \; \qform[\xi] \; \leq C \;  \|\xi\|_{\hcfpiu}\,,
\eeq
where $c, C$ are two positive constants. Estimate  \eqref{cC}  implies that  $\qform[\xi]$ is finite  for any  $\xi \in \dom(\qform)$ and therefore our definition of the quadratic forms \eqref{form} and \eqref{qform} is well-posed. 
On the other hand, if in addition we assume $\Lambda(m,N) <1$, from  \eqref{cC}  we conclude that $\qform$ defines a norm equivalent to the $H^{1/2}$-norm. This is the crucial ingredient for the proof of Theorem \ref{clbou}.

\begin{proof}[Proof of Theorem \ref{clbou}]  From \eqref{form} and \eqref{fido} we have
\bml{
\form [\psi] \geq -\lambda \|\psi\|^2_{L^2_{\mathrm{f}}} + N \qform [\xi] \geq -\lambda \|\psi\|^2_{L^2_{\mathrm{f}}}	\\
 + N \!\! \int_{\bR^{3N-3}} \!\!\!\! \! \diff \kv_1 \cdots \diff \kv_{N-1} \: \big| \hat\xi(\kv_1, \ldots, \kv_{N-1}) \big|^2 \bigg[ \al + 2 \pi^2  \Big(\! 1 - \Lambda(m,N) \!\Big)  \bigg(\! \frac{m}{m\!+\!1} \sum_{i = 1}^{N-1}\! k_i^2 + \la \! \bigg)^{\!\!1/2}  \bigg] 
}
for any $\psi \in \dom(\form)$ and $\lambda >0$. Therefore, if $\alpha \geq 0$ the form $\form$ is positive and if $\alpha<0$ the lower bound \eqref{inff} holds. Let us now prove that $\form$ is closed. We choose 
\beq
\lambda >
	\begin{cases} 
		0	& \text{if} \;\; \alpha \geq 0, 	\\
		\f{\alpha^2}{4 \pi^4 \big(1-\Lambda(m,N)\big) }	&	\text{if} \;\; \alpha <0\,,
	\end{cases}
\eeq
and consider the form $\form^{\lambda}[\psi] :=\form [\psi] +\lambda \|\psi\|^2_{L^2_{\mathrm{f}}}$ defined on $\dom(\form)$. Let $\{\psi_n \}$ be a sequence in $\dom(\form)$ such that
\beq
\lim_{n \rightarrow \infty} \|\psi_n - \psi \|_{L^2_{\mathrm{f}}} =0\,, \hspace{1cm}	\lim_{n,m \rightarrow \infty} \form^{\lambda}[\psi_n - \psi_m] =0\,,
\eeq
where $\psi \in L^2_{\mathrm{f}}(\bR^{3N})$. From the definition of $\dom(\form)$ and $\form^{\lambda}$  (see \eqref{form}, \eqref{form domain}) we have $\psi_n = \phi_n^{\lambda} + \pot \xi_n$, with $\phi_n^{\lambda} \in H^1_{\mathrm{f}}(\bR^{3N})$, $\xi_n \in H^{1/2}_{\mathrm{f}}(\bR^{3N-3})$, and 
\beq
\form^{\lambda}[\psi_n - \psi_m]= \mathcal{F}_0[\phi_n^{\lambda} - \phi_m^{\lambda}] + N \qform[\xi_n -\xi_m]\,.
\eeq
This, together with \eqref{cC}, implies that $\{\phi_n^{\lambda} \}$ is a Cauchy sequence in $ H^1_{\mathrm{f}}(\bR^{3N})$ and $\{ \xi_n\}$ is a Cauchy sequence in $H^{1/2}_{\mathrm{f}}(\bR^{3N-3})$. Let us denote by $\phi^{\lambda}$ and $\xi$ the corresponding limits. From the explicit expression of the potential \eqref{pot xi} we notice that 
\beq
\| \pot \xi \|_{L^2_{\mathrm{f}}} \leq c \, \|\xi\|_{L^2_{\mathrm{f}}}\,,
\eeq
where $c>0$. Hence,
\beq
\lim_{n \rightarrow \infty} \|\psi_n - (\phi^{\lambda} + \pot \xi) \|_{L^2_{\mathrm{f}}}= \lim_{n \rightarrow \infty}  \| (\phi_n^{\lambda} - \phi^{\lambda}) + \pot ( \xi_n - \xi) \|_{L^2_{\mathrm{f}}}       =0\,.
\eeq
Since the limit of the $\psi_n$'s is unique, $\psi = \phi^{\lambda} + \pot \xi$. Therefore $\psi \in \dom(\form)$ and $\lim_{n \rightarrow \infty} \form^{\lambda}[\psi-\psi_n]=0$. This shows that the form $\form^{\lambda}$ is closed and a fortiori $\form$ is.
\end{proof}


\section{Unboundedness from Below of $ \form $} 
\label{instability: sec}
		
This section is devoted to the proof of Theorem \ref{ub1}. 
As we shall see, what makes an instability condition hard to prove is the restriction to antisymmetric wave functions. 

In fact, the proof relies on the explicit evaluation of the charge form $ \qform $ on a trial function, i.e., a convenient sequence of charges with energy going to $ -\infty $. Identifying one such sequence is easy when $ N = 2 $ because $ \qform $ is in practice the same as the reduced form $ F_{1} $ -- see \eqref{michelebuliccio} below -- 
and the analysis performed in Section \ref{stability: sec} suggests that a convenient $ Q_n(\pv) $ has to be chosen 
in the subspace with angular momentum $ \ell=1 $ and such that in the position representation it becomes peaked at the origin as $ n \to \infty $ (i.e., two identical fermions coming arbitrarily close).

When $ N>2 $, on the other hand, a natural trial function satisfying the antisymmetry constraint would be the Slater determinant of $N$ one-particle charges, one of which is $ Q_n $ itself. A convenient choice is driven by the physical idea of a $N$-body configuration that contains precisely the $(2\!+\!1)$-body structure minimizing the energy with $ N = 2 $, whereas all remaining particles are placed far away in space so that there is no or negligible interference with the two-body state. This results in a $N$-particle Slater determinant  between $ Q_n $ and $N-1$ copy of a different component (see \eqref{min sequence} below). The fermionic character of the trial function is thus fulfilled by construction and optimising the choice of the second component produces only higher order symmetry correlations.

	Throughout this section we assume that $ \la $ is a positive number such that $ C_1 \leq \la^{-1} \leq C_2 $ for two finite constants $ C_1, C_2 < \infty $, which in particular will allow us to incorporate error factors proportional to $ \la^{-1} $ into a constant $ C $.
 	
	\begin{proof}[Proof of Theorem \ref{ub1}]
		In order to prove instability of the form $ \form $, it is enough to produce a sequence of normalised charges $ \xi_n \in H_{\mathrm{f}}^{1/2}(\R^{3(N-1)}) $ such that
		\beq
			\lim_{n \to \infty} \qform[\xi_n] = - \infty,
		\eeq
		since the sequence of states $ \pot \xi_n $ then satisfies
		\beq
			\form[\pot \xi_n] = - \la \lf\| \pot \xi_n \ri\|_{L^2(\R^{3N})}^2 + N \qform[\xi_n] \xrightarrow[\;n\to\infty\;]{} - \infty.
		\eeq
		\underline{Case $ N = 2 $}. The result was already proved in \cite{FT}, but we repeat here the argument for we use a slightly different trial function that turns out to be useful in the general case $ N > 2 $. Owing to  \eqref{qform 1}, 
		\beq\label{michelebuliccio}
			\qform[\xi] -  \alpha \| \xi \|_{L^2(\R^{3})}^2 = \sqrt{\la} F_1 \lf[ Q \ri]
		\eeq
		where $ Q(\pv) = \la^{3/4} \xi(\sqrt{\la}\pv) $ (recall \eqref{charge Qk}). Then we only need to produce $ Q_n \in L^2(\RT) $ such that $ \lim_{n \to \infty} F_1[Q_n] = - \infty $. Note that no constraint is imposed on the symmetry properties of $ Q_n $. In fact, according to the discussion of Section \ref{stability: sec} (see  \eqref{formula0}, \eqref{formula1} and \eqref{S10}), we can take each $ Q_n $ in the subspace with angular momentum $ l = 1 $ and such that the support of its $\sharp-$transform defined in \eqref{formula2} gets concentrated at the origin. Explicitly, we choose
		\beq
			\label{qnga}
			\qnga(\kv) : = n^{-3/2} \qga(n^{-1} k) Y_1^{0}(\vartheta_k),
		\eeq
				\beq
			\label{Qgamma}
			\qga(p) : =  \pi^{-1/4} c_{\gamma} \gamma^{1/2} p^{-1} \exp \lf\{ - \hbox{$\frac{1}{8\gamma^{2} }$} \ri\} \exp \lf\{ - \half \gamma^2 \lf( \log p \ri)^2 \ri\} \Theta(p-1),
		\eeq
		where $ \gamma \in (0, 1) $ is a variational parameter, $ \Theta $ is the Heaviside function, i.e., $ \Theta(p) = 1 $ if $ p \geq0 $ and $ 0 $ otherwise, and $ c_{\gamma} $ is a normalisation constant. 
		By direct computation,
		\beq
			\lf\| \qnga \ri\|^2_{L^2} = \int_1^{\infty} \diff p \: p^2 \lf| \qga(p) \ri|^2 = \tx\frac{c^2_{\gamma}}{\sqrt{\pi}} \disp\int_{0}^{\infty} \diff t \: \exp\lf\{ - t^2 + \tx\frac{t}{\gamma} - \tx\frac{1}{4\gamma^2} \ri\} = \half c^2_{\gamma} \: \lf( 1 + \mathrm{erf}\lf\{ \tx\frac{1}{2\gamma} \ri\} \ri).
		\eeq
		Imposing $\qnga$ to be normalised yields
		\beq
			\label{cgamma}
			1 \leq c^2_{\gamma} = 2 \lf[ 1 + \mathrm{erf}\lf\{ \tx\frac{1}{2\gamma} \ri\} \ri]^{-1} \leq 1 + C\gamma \exp\lf\{-\tx\frac{1}{4\gamma^2} \ri\}.
		\eeq 
(see, e.g., \cite[Eq. (7.1.13)]{AS}). It is also useful to compute the following integral for  $ a \geq -1 $:
		\bml{
			\label{qga est}
			\int_0^{\infty} \diff p \: p^{2+a} \lf| \qga(p) \ri|^2 = \tx{\frac{1}{\sqrt{\pi}}} \gamma c_{\gamma}^2 \exp \lf\{ -\tx{\frac{1}{4\gamma^2}} \ri\} \disp\int_{0}^{\infty} \diff t \: \exp \lf\{ - \gamma^2 t^2 + (1+a) t \ri\}	\\
			 =\half c_{\gamma}^2 \lf( 1 + \mathrm{erf} \lf\{\tx\frac{1+a}{2\gamma} \ri\} \ri) \exp \lf\{ \tx{\frac{1}{4\gamma^2}} (2a+a^2) \ri\} \leq  \lf( 1 + C \gamma \ri) \exp \lf\{ \tx{\frac{1}{4\gamma^2}} (2a+a^2) \ri\},	
		}
	where we used
		\bdm
			\half c_{\gamma}^2 \lf( 1 + \mathrm{erf} \lf\{\tx\frac{1+a}{2\gamma} \ri\} \ri) = \frac{ 1 + \mathrm{erf} \lf\{\tx\frac{1+a}{2\gamma} \ri\}}{ 1 + \mathrm{erf} \lf\{\tx\frac{1}{2\gamma} \ri\}} \leq 1 + \frac{2}{\sqrt{\pi}} \int_{\frac{1}{2\gamma}}^{\frac{1+a}{2\gamma}} \diff t \: e^{-t^2} \leq 1 + C \gamma^{-1} \exp\lf\{- \tx\frac{1}{4 \gamma^2} \ri\} \leq 1 + \OO(\gamma).
		\edm
		Using the decomposition \eqref{Fzl}, as well as the scaling law of $ \qnga $ with $n$, we have
		\beq
			\label{est F1 1}
			F_1[\qnga] = G_{1}^{\mathrm{diag}}\lf[n^{-3/2} \qga(n^{-1} p)\ri] + G_{1,1}^{\mathrm{off}}\lf[n^{-3/2} \qga(n^{-1} p)\ri] = n \lf[ G_{n^{-2}}^{\mathrm{diag}}\lf[\qga\ri] + G_{n^{-2},1}^{\mathrm{off}}\lf[\qga\ri] \ri].
		\eeq
We estimate the diagonal term in \eqref{est F1 1} as
\begin{equation}\label{approx Goff}
\begin{split}
G_{n^{-2}}^{\mathrm{diag}}\lf[\qga\ri]\;&\leq\;2 \pi^2 \frac{\sqrt{m(m+2)}}{m+1} \lf(1 + \OO(n^{-1}) \ri) \big\| Q_{\gamma}^{\sharp} \big\|_{L^2}^2	\\
&\leq\;2 \pi^2 \frac{\sqrt{m(m+2)}}{m+1} \exp \lf\{ \tx\frac{3}{4\gamma^2} \ri\} \lf[1 + \OO(n^{-1}) + \OO(\gamma) \ri],
\end{split}
\end{equation}
where we used
		\beq
			\label{tilde Q norm}
			\big\| Q^{\sharp}_{\gamma} \big\|_{L^2}^2 = \int_{\R} \diff k \: e^{4k} \: \big| \hat{Q}_{\gamma}\big(e^k\big) \big|^2 = \int_{0}^{\infty} \diff p \: p^3 \: | Q_{\gamma}(p) |^2 \leq  (1 + C \gamma ) \exp \lf\{ \tx{\frac{3}{4\gamma^2}} \ri\}.
		\eeq
		As for the off-diagonal term in \eqref{est F1 1},
\beq\label{Goff1}
G^{\mathrm{off}}_{n^{-2},1} \lf[ \qga \ri] =   G^{\mathrm{off}}_{0,1} \lf[ \qga \ri] + \mathcal{R},
\eeq
where
\bml{
 			|\mathcal{R}| \leq C \int_{\R^3} \diff \sv \int_{\R^3} \diff \tv \: \bigg| \lf(s^2 + t^2 + \tx\frac{2}{m+1} \sv \cdot \tv + n^{-2}\ri)^{-1} - \lf(s^2 + t^2 + \tx\frac{2}{m+1} \sv \cdot \tv\ri)^{-1} \bigg| \lf| \qga(s) \ri| \lf| \qga(t) \ri|	\\
			\leq C n^{-2} \int_{\R^3} \diff \sv \int_{\R^3} \diff \tv \: \frac{\lf| \qga(s) \ri| \lf| \qga(t) \ri|}{(s^2 + t^2)^{2}}  \leq C n^{-2} \int_{\R^3} \diff \sv \: \lf| \qga(s) \ri|^2 \int_{1}^{\infty} \diff t \: t^{-2} \leq \OO(n^{-2}). \label{R1}
		}
Moreover,
\begin{equation}\label{Goff est}
\begin{split}
G^{\mathrm{off}}_{0,1} \lf[ \qga \ri]\;&=\;\int_{\R} \diff k \: S_1(k) \big| Q^{\sharp}_{\gamma}(k) \big|^2  \\
& \leq\; S_1(0) \big\| Q^{\sharp}_{\gamma} \big\|_{L^2}^2 + \int_{\R} \diff k \: \lf( S_1(k) - S_1(0) \ri) \big| Q^{\sharp}_{\gamma}(k) \big|^2 \\
&\leq\; S_1(0)  \exp \lf\{ \tx{\frac{3}{4\gamma^2}} \ri\} \lf(1 + \OO(\gamma)  \ri) + C \int_{\R} \diff k \: \sqrt{|k|} \big| Q^{\sharp}_{\gamma}(k) \big|^2 \\
&=\; - 2 \pi^2 \frac{\sqrt{m(m+2)}}{m+1} \Lambda(m,2) \exp \lf\{ \tx\frac{3}{4\gamma^2} \ri\} \lf( 1 + \OO(\gamma)  \ri) + C \int_{\R} \diff k \: \sqrt{|k|} \big| Q^{\sharp}_{\gamma}(k) \big|^2
\end{split}
\end{equation}
where we used \eqref{S10} and the elementary estimate $S_1(k)-S_1(0) \leq C \sqrt{|k|}$. To estimate the last integral in \eqref{Goff est} we observe that
\begin{equation}\label{Qshes}
\begin{split}
Q^{\sharp}_{\gamma}(k)\;&=\;\pi^{-1/4} c_{\gamma} \gamma^{1/2}  \exp\lf\{\tx-\f{1}{8 \gamma^2} \ri\}    \f{1}{\sqrt{2\pi}} \int_0^{\infty} \!\!\!dx\, 
\exp\lf\{\tx-ikx -\f{\gamma^2}{2} x^2 +x\ri\}  \\
&=\;\pi^{-1/4} c_{\gamma} \gamma^{-1/2} 
\exp\lf\{\tx-\f{1}{8 \gamma^2}\ri\} \bigg(
\exp\lf\{\tx-\f{k^2}{2\gamma^2} -i\f{k}{\gamma^2} +\f{1}{2\gamma^2}\ri\} - \f{1}{2 \sqrt{\pi} z}(1+r(z)) \bigg),
\end{split}
\end{equation}
where
\beq
z= \f{1-ik}{\sqrt{2}\, \gamma}\,, \hspace{1cm} |r(z)|\leq \f{\gamma^2}{\sqrt{1+k^2}}\,.
\eeq
Therefore,
\begin{equation}\label{sqrQ}
\begin{split}
\!\!\!\!\!\!\int_{\bR}\!\!\!dk\, \sqrt{|k|} |Q^{\sharp}_{\gamma} (k)|^2 \;&\leq\;\f{2 \, c_{\gamma}^2}{\sqrt{\pi} \gamma}  
\exp\lf\{\tx\f{3}{4\gamma^2}\ri\} \int_{\bR}\!\!\! dk\, \sqrt{|k|} \, 
\exp\lf\{\tx-\f{k^2}{\gamma^2}\ri\}  
+ \f{4 \, c_{\gamma}^2 \, \gamma}{\pi^{3/2}} 
\exp\lf\{\tx-\f{1}{4\gamma^2} \ri\}
\int_{\bR} \!\!\!dk \, \f{\sqrt{|k|}}{1+\, k^2} \\
& \leq\;C \, \sqrt{\gamma} \; \exp\lf\{\tx\f{3}{4 \gamma^2} \ri\} \Big( 1 + \sqrt{\gamma} \, 
\exp\lf\{\tx-\f{1}{\gamma^2}\ri\} \Big).
\end{split}
\end{equation}
Using \eqref{approx Goff}, \eqref{Goff1}, \eqref{R1}, \eqref{Goff est}, \eqref{sqrQ} in  \eqref{est F1 1} we finally obtain
		\beq
			F_1[\qnga] \leq 2 \pi^2 \frac{\sqrt{m(m+2)}}{m+1} n \exp \lf\{ \tx\frac{3}{4\gamma^2} \ri\} \lf[ 1 - \Lambda(m,2) + \OO(\sqrt{\gamma}) + \OO(n^{-1}) \ri] \xrightarrow[\;n\to\infty\;]{} - \infty,
		\eeq
		if $ \Lambda(m,2) > 1 $ and $ \gamma $ is taken small enough (independent of $ n $).

\noindent \underline{Case $N>2$}. 
		As mentioned at the beginning of this section, this case is more complicated, for the trial sequence $ \xi_n $ must be antisymmetric under the exchange of any variable, i.e., $\xi_n \in \ldf(\R^{3N-3}) $, and at the same time we want $ \hat\xi_n(\kv_1,\ldots,\kv_{N-1}) $ to behave like $ \qnga(\kv_1) $ once the other degrees of freedom are traced out. Looking for $\xi_n$ matching these two requirements is an example of the well-known representability problem (see, e.g., \cite{LS}), i.e., the search for sufficient conditions to impose on a one-particle density matrix so that it can be obtained as the reduced  density matrix of a fermionic many-body state. We remark that the solution is known only in some special cases and is non-trivial. Our choice here is a trial state that is as close as possible to an uncorrelated state, which is given by an antisymmetric wave function containing $ \qnga $. Explicitly,
		\beq
 			\label{min sequence}
			\hat\xi_n(\kv_1, \ldots, \kv_{N-1}) : = \frac{1}{\sqrt{(N-1)!}}
			\lf|
			\begin{array}{ccccc}
				\qnga(\kv_1)	&	\Xi_{\beta,2}(\kv_1)	&	\cdots 	&	\Xi_{\beta,N-1}(\kv_{1})		\\
				\qnga(\kv_2)	&	\Xi_{\beta,2}(\kv_2)	&	\cdots 	&	\Xi_{\beta,N-1}(\kv_{2})		\\
				\vdots 	&	\vdots 	&	\mbox{} 	&	\vdots		\\
				\qnga(\kv_{N-1})	&	\Xi_{\beta,2}(\kv_{N-1})	&	\cdots 	&	\Xi_{\beta,N-1}(\kv_{N-1})	
			\end{array}
			\ri|\,,
		\eeq
		where $ \qnga $ is defined in \eqref{qnga}, $ 0 < \beta \ll 1 $ is another variational parameter,
		\beq
			\label{chibm}
			\chibm(\kv) := (4\pi)^{-1/2} \beta^{-3/2} \: \Xi(\beta^{-1} k) \: \exp\lf\{ i l \varphi_k \ri\},
		\eeq
		$ l \in \N $, $ \Xi \in C^{\infty}_0(\R^+) $ is real-valued, with support in $ (0,1) $, and such that
		\beq
			\label{G normalized}
			\int_{0}^1 \diff k \: k^2 \: \Xi^2(k) = 1.
		\eeq
		Note that, since the two functions $ \qnga $ and $ \chibm $, $ l > 0 $, are orthonormal by construction, the function \eqref{min sequence} belongs to $ \ldf(\R^{3(N-1)}) $ and  is normalised. Moreover, the supports of $ \qga $ and $ \Xi $ do not intersect, which implies that the supports of $ \qnga $ and $ \chibm $ are disjoint as well, provided $ \beta \leq n $, which follows from the assumptions on $ \beta $. 
		
		We can now evaluate $\dqform[\xi_n]$. We start by estimating the diagonal part. Using the exchange symmetry and the definition of $ L_{\la} $ in \eqref{Lla}),
		\bml{
 			\label{energy est 0}
			\dqform[\xi_n] = \al +  \frac{1}{(N-2)!} \int_{\R^{3(N-1)}} \diff \kv_1 \diff \kkv \: L_{\la}(\kv_1,\ldots,\kv_{N-1}) \lf| \qnga(\kv_1) \ri|^2 	\cdot \\
			\sum_{\sigma,\tau \in \mathcal{P}_{N-1}} \prod_{l,j =2}^{N-1} \sgn(\sigma) \sgn(\tau) \Xi_{\beta,l}^*(\kv_{\sigma(l)}) \Xi_{\beta,j}(\kv_{\tau(j)}),
			}
		where $ \mathcal{P}_{N-1} $ is the group of permutations of $ N-2 $ elements $2, \ldots, N - 1 $ and $ \sgn(\sigma) $ denotes the sign of any $ \sigma \in \mathcal{P}_N $. All the other terms vanish because of the integral of the product $ \qnga(\kv_i) \chib(\kv_i) $, which is pointwise zero thanks to the disjoint supports of the functions. Extracting the main factor
		\bdm
			\sqrt{\tx\frac{m(m+2)}{(m+1)^2} k_1^2 + \la}\,,
		\edm
		and bounding the rest by means of the inequality
		\beq
			\label{useful ineq 1}
			\sqrt{a + b} \leq \sqrt{|a|} + \sqrt{|b|},	\hspace{1cm}	\mbox{for} \:\: a + b \geq 0\,,	
		\eeq
		we obtain 
\begin{equation}\label{Lla estimate}
\begin{split}
L_{\la}(\kv_1, \ldots, \kv_{N-1}) \;&\leq\; 2 \pi^2 \sqrt{\tx\frac{m(m+2)}{(m+1)^2} k_1^2 + \la} \,\bigg\{ 1 + \lf(\tx\frac{m(m+2)}{(m+1)^2} k_1^2 + \la\ri)^{-1/2}  \times \\
& \qquad\qquad\times\bigg[ \tx\frac{m(m+2)}{(m+1)^2} \disp\sum_{i = 2}^{N-1} k_i^2 + \tx\frac{2m}{(m+1)^2} \bigg| \disp\sum_{j > 1} \kv_1 \cdot \kv_j +  \disp\sum_{1 < i <j} \kv_i \cdot \kv_j \bigg| \bigg]^{1/2} \bigg\}	\\
&\leq\;2 \pi^2 \sqrt{\tx\frac{m(m+2)}{(m+1)^2} k_1^2 + \la} \bigg\{ 1 + C_{N} \disp\sum_{i = 2}^{N-1} \lf( k_i + \sqrt{k_1k_i} \ri) \bigg\}\,.
\end{split}
\end{equation}
The diagonal term can be estimated as	
		\bml{
 			\label{energy est 1}
			\dqform[\xi_n] - \al \leq  2 \pi^2\int_{\R^{3(N-1)}} \diff \kv_1 \cdots \diff \kv_{N-1} \:  \sqrt{\tx\frac{m(m+2)}{(m+1)^2} k_1^2 + \la} \lf| \qnga(\kv_1) \ri|^2 \prod_{l =2 }^{N-1} \lf| \Xi_{\beta,0}(\kv_l)\ri|^2	\\
			+ C_N \int_{\R^{3(N-1)}} \diff \kv_1 \cdots \diff \kv_{N-1} \: \lf|L_{\la}(\kv_1, \kkv) - 2 \pi^2 \sqrt{\tx\frac{m(m+2)}{(m+1)^2} k_1^2 + \la} \ri|  \lf| \qnga(\kv_1) \ri|^2 \prod_{l =2 }^{N-1} \lf| \Xi_{\beta,0}(\kv_l)\ri|^2,
		}
		thanks to the orthogonality of functions $ \Xi_{\beta,l} $  and  $ \Xi_{\beta,l^{\prime}} $ for $ l \neq l^{\prime} $. Thus, by \eqref{Lla estimate}, $ \dqform[\xi_n] - \al $ is bounded from above by
\begin{equation}
\begin{split}
\!\!\!\!\!\!\!2 \pi^2\!\!\!& \int_{\R^3} \diff \kv_1 \sqrt{\tx\frac{m(m+2)}{(m+1)^2} k_1^2 + \la} \lf| \qnga(\kv_1) \ri|^2 \int_{\R^{3(N-2)}} \diff \kkv \bigg[ 1 + C_N \bigg( k_2 + \sqrt{k_1 k_2} \bigg) \bigg] \prod_{l =2 }^{N-1} \lf| \Xi_{\beta,0}(\kv_l)\ri|^2 \\
&\leq\;2 \pi^2 n \int_{\R^3} \diff \kv_1  \sqrt{\tx\frac{m(m+2)}{(m+1)^2} k_1^2 + \frac{\la}{n^2}}  \lf| \qga(\kv_1) \ri|^2 \int_{0}^{1} \diff k_2 \: k_2^2\lf[ 1 + C_N \lf( \beta k_2 + \sqrt{n \beta k_1 k_2} \ri) \ri] \lf| \Xi(k_2) \ri|^2 \\
& \leq 2 \pi^2 n \tx\frac{\sqrt{m(m+2)}}{m+1} \lf(1 + \OO(n^{-1}) \ri) \disp\int_{1}^{\infty} \diff k_1 \: k_1^3 \lf[ 1 + C_N \lf( \beta + \sqrt{n \beta k_1} \ri) \ri] \lf| \qga(k_1) \ri|^2 \\
& \leq\;2 \pi^2 n \tx\frac{\sqrt{m(m+2)}}{m+1}  \exp \lf\{ \tx\frac{3}{4\gamma^2} \ri\}  \lf[1 + C _N \lf( \sqrt{n \beta} \exp \lf\{ \tx{\frac{9}{16 \gamma^2}} \ri\} + \gamma +  \beta + n^{-1} \ri) \ri]
\end{split}
\end{equation}
where we used \eqref{qga est}. We now compute the off-diagonal term (recall \eqref{oqform}).  Owing to the exchange symmetry, the pre-factor $ N-1 $ cancels with the normalisation factor of $ \xi_n $ and
		\bml{
 			\label{energy est 3}
			\oqform[\xi_n] = \int_{\R^{3N}} \diff \sv \diff \tv \diff \kkv \:	\green(\sv, \tv, \kkv) \bigg\{ \qnga^*(\sv) \qnga(\tv) \times 	\\
			\times\frac{1}{(N-2)!}\sum_{\sigma,\tau \in \mathcal{P}_{N-1}} \prod_{l,j =2}^{N-1} \sgn(\sigma) \sgn(\tau) \Xi_{\beta,l}^*(\kv_{\sigma(l)}) \Xi_{\beta,j}(\kv_{\tau(j)}) \bigg\}  + \mathcal{R}
		}
		where $ \mathcal{R} $ contains some remainder terms. We estimate the leading term (first term on the r.h.s.~of \eqref{energy est 3}) from above by
		\bml{
 			\label{energy est 4}
			n \int_{\R^{6}} \diff \sv \diff \tv \:  G_{\la/n^2}(\sv, \tv) \qga(s) \qga(t) Y_1^0(\vartheta_s) Y_1^0(\vartheta_t)	\\
			 + C_N \int_{\R^{3N}} \diff \sv \diff \tv \diff \kkv \:	\green(\sv, \tv, \kkv) \green(\sv, \tv) \lf[ k_2^2 + (s+t) k_2 \ri] 	\lf| \qnga(\sv) \ri| \lf| \qnga(\tv) \ri| \prod_{l =2 }^{N-1} \lf| \Xi_{\beta,0}(\kv_l)\ri|^2,			}
		where in the first term we replaced $ \green(\sv, \tv, \kv_2, \ldots, \kv_{N-1}) $ with $ \green(\sv, \tv, \zerov, \ldots, \zerov) = :  \green(\sv, \tv) $ and we exploited the orthogonality of functions $ \Xi_{\beta,l} $  and  $ \Xi_{\beta, l^{\prime}} $ for $ l \neq l^{\prime} $.
		The first term in the expression above was bounded  in \eqref{Goff est}.  Using
		\beq
			\label{kernel est 1}
			\green(\sv,\tv,\kv_2, \ldots, \kv_{N-1}) \leq \bigg[ \frac{m}{m+1} \big(s^2 + t^2 
			\big) + \la \bigg]^{-1}
		\eeq 
	and the elementary inequality $x+y \leq \sqrt{x^2 +1} \sqrt{y^2 +1}$, $x,y \geq 0$, 	the second term in \eqref{energy est 4} can be estimated as
\begin{equation}\label{off2}
\begin{split}
C_N \int_{\R^{3N}}& \diff \sv \diff \tv \diff \kkv   \f{k_2^2 +(s+t) k_2}{\big[ s^2 +t^2 + \lambda \f{m+1}{m} \big]^2}  \lf| \qnga(\sv) \ri| \lf| \qnga(\tv) \ri| \prod_{l =2 }^{N-1} \lf| \Xi_{\beta,0}(\kv_l)\ri|^2 \\
&\leq\;C_N \, n^{-1}  \int_{\R^{6}} \!\! \diff \sv \diff \tv  \f{\beta^2 +n \beta (s+t) }{\big[ s^2 +t^2 + \lambda \f{m+1}{m} n^{-2} \big]^2}  \lf| \qga(\sv) \ri| \lf| \qga(\tv) \ri|\\
&\leq\; C_N n^{-1} (\beta^2 + n \beta)  \bigg[ \sup_{\tv} \int_{\R^{3}} \!\! \diff \sv   \f{1 }{\big[ s^2 +t^2 + \lambda \f{m+1}{m} n^{-2} \big]^2} \bigg] \int_{\R^3} \!\! \diff \pv \, (p^2 +1) |Q_{\gamma}(\pv)|^2 \\
&\leq C_N (\beta^2 + n \beta) (1+\gamma) \exp \lf\{ \tx\frac{2}{\gamma^2} \ri\}\,.
\end{split}
\end{equation}
The rest $ \mathcal{R} $ in \eqref{energy est 3} contains several terms but it is not difficult to see that most of them vanish because of the disjoint supports of $ \qnga $ and $ \chib $. What remains is
\begin{equation}\label{remainder 1}
\begin{split}
[(N-3)!]^{-1} &\int_{\R^{3N}} \diff \sv \diff \tv \diff \kv_2 \cdots \diff \kv_{N-1} \:	\green(\sv, \tv, \kv_2, \ldots, \kv_{N-1}) \: \Xi_{\beta,1}^*(\sv) \Xi_{\beta,1}(\tv)  \lf| \qnga(\kv_2) \ri|^2 \times \\
& \qquad\qquad\times \sum_{\sigma,\tau \in \mathcal{P}_{N-2}} \prod_{l,j = 3}^{N-1} \sgn(\sigma) \sgn(\tau) \Xi_{\beta,l}^*(\kv_{\sigma(l)}) \Xi_{\beta,j} (\kv_{\tau(j)}) \\ 
&\leq\;	C_N \int_{\R^{6}} \diff \sv \diff \tv \:  \lf| \Xi_{\beta,1}(\sv)  \ri| \lf| \Xi_{\beta,1}(\tv) \ri| \leq C_N \beta^3,
\end{split}
\end{equation}
where we exploited the exchange symmetry again, as well as the properties of $ \Xi $ (in particular $ \supp(\Xi) \subset (0,1) $) and \eqref{kernel est 1}. It is understood that the sum over permutations as well as the following factor in \eqref{remainder 1} is not present when $N = 3 $. Putting together \eqref{energy est 3}, \eqref{energy est 4}, \eqref{off2} and \eqref{remainder 1}, we obtain 
		\beq
 			\label{energy est 8}
			\oqform[\xi_n] \leq n G^{\mathrm{off}}_{\la/n^{2},1} \lf[ \qga \ri] + C_N \lf(\beta^2 + n \beta \ri) \exp \lf\{ \tx\frac{2}{\gamma^2} \ri\},
		\eeq
		and finally
		\bml{
 			\label{energy est final}
			\qform[\xi_n] \leq 2 \pi^2 n \frac{\sqrt{m(m+2)}}{m+1} \exp\lf\{ \tx\frac{3}{4\gamma^2} \ri\} \Big\{ 1 -  \Lambda(m,2) 	\\
			   + C_N \lf[ \al \,  n^{-1} + \sqrt{\gamma} + \sqrt{n \beta} \exp \lf\{ \tx{\frac{9}{16 \gamma^2}} \ri\} + (n^{-1} \beta^2 +\beta) \exp\lf\{ \tx\frac{5}{4\gamma^2} \ri\}  \ri] \Big\}\,.
		}
		By assumption, $  1 -  \Lambda(m,2) < 0 $ and we choose $ \beta \ll n^{-1} $ as, say, $ \beta = n^{-2} $. Thus, we can always find some small $ \gamma = \OO(1) > 0 $, such that 
		\beq
			\qform[\xi_n] 
			\xrightarrow[\;n\to\infty\;]{} - \infty
		\eeq
which concludes the proof.
\end{proof}

\vspace{1cm}
\n		
{\bf Acknowledgments}. M.C. acknowledges the support of the European Research Council under the European Community Seventh Framework Program (FP7/2007-2013 Grant Agreement CoMBos No. 239694).

\vs

\section*{Appendix}

\renewcommand{\theequation}{A.\arabic{equation}}
\setcounter{equation}{0}
\setcounter{subsection}{0}
\setcounter{pro}{0}
\renewcommand{\thesection}{A}


\label{Appendix A}

Here we describe the formal procedure for the construction of the quadratic form $\form$.   
We start  from the  Hamiltonian \eqref{cm Hamiltonian} written in the Fourier space
\bml{
(\widehat{H \psi})(\kv_1,\ldots, \kv_N) = \bigg(\sum_{i=1}^N k_i^2 +\f{2}{m\!+\!1}\sum_{i<j} \kv_i \cdot \kv_j \bigg) \hat{\psi}(\kv_1,\ldots, \kv_N) \\
  +\f{\mu}{(2 \pi)^3} \int_{\bR^3} \!\! \diff \sv \,\hat{\psi} (\kv_1,\ldots, \kv_{i-1},\sv,\kv_{i+1},\ldots,\kv_N)
}
and we consider the corresponding quadratic form, regularized by means of an ultra-violet cut-off
\bml{
	\label{ren Form}
	\renform[\psi] : = \int_{\R^{3N}} \diff \kv_1 \cdots \diff \kv_N \bigg\{ \sum_{i =1}^N k_i^2 + \f{2}{(m+1)} \sum_{i < j} \kv_i \cdot \kv_j \bigg\} \lf| \hat{\psi}(\kv_1, \ldots, \kv_N) \ri|^2	\\
	 + \frac{\mu(\alpha,R)}{(2\pi)^{3}} \sum_{i = 1}^N   \int_{\R^{3N}} \diff \kv_1 \cdots \diff \kv_N \: \chi_R(k_i)  \hat{\psi}^*(\kv_1,\ldots, \kv_N) \times 	\\
	\times\int_{\RT} \diff \sv \: \chi_R(s) \hat{\psi}(\kv_1,\ldots, \kv_{i-1},\sv,\kv_{i+1},\ldots,\kv_N)\,.
}
Here $ \psi \in H^1(\R^{3N}) $, $ \chi_R(s) $ is the characteristic function of the three-dimensional ball $ s \leq R $, and $ \alpha $ is a  parameter that has the role of a renormalised coupling constant.  Note also that we introduced in $ \mu$ a dependence on $ R $: the choice of such an explicit dependence will be the main content of the renormalisation procedure.

We now define the ``surface charges'' $ \xi_{i}^R \in L^2(\R^{3N-3}) $ associated with $ \psi \in H^1(\R^{3N}) $ as
\beq
	\label{ren charges}
	\hat{\xi}_i^R(\kv_1,\ldots,\kv_{N-1}) : = \frac{\mu(\alpha,R)}{(2\pi)^3} \int_{\RT} \diff \sv \: \chi_R(s) \hat{\psi}(\kv_1,\ldots,\kv_{i-1},\sv,\kv_{i}, \ldots, \kv_{N-1}),
\eeq
and the corresponding ``volume charges'' $\hat{\rho}_i^R (\kv_1,\ldots,\kv_N):= \chi_R(k_i)  \: \hat{\xi}_i^R(\breve \kkv_i)$. Further, we introduce    the  ``potential'' 
\beq
	\label{ren potential}
	\widehat{\pot  \rho^R}(\kv_1, \ldots, \kv_N) : = \sum_{i=1}^N \green(\kv_1,\ldots,\kv_N) \: \chi_R(k_i)  \: \hat{\xi}_i^R(\breve \kkv_i),
\eeq
where $G_{\la}$ is defined in \eqref{Green function} for any $ \la > 0 $.
Setting
\beq
	\label{ren regular part}
	\hat{\phi}_{\la}^R : = \hat{\psi} - \widehat{\pot \rho^R},
\eeq
we have
\beq
	\label{ren form decomposition}
 	\renform[\psi] = \F_0\lf[\phi_{\la}^R\ri] + \la \lf\| \phi_{\la}^R \ri\|_{L^2(\R^{3N})}^2 - \la \lf\| \psi \ri\|_{L^2(\R^{3N})}^2 + \renqform\lf[\xi^R\ri],
\eeq
with $ \F_0[\phi] : = \bra{\phi} H_0 \ket{\phi} $, and
\bml{
	\label{ren qform}
	\renqform\lf[\xi\ri] : = - \sum_{i=1}^N \int_{\R^{3N}} \diff \kv_1 \cdots \diff \kv_N \: \chi_R(k_i) \: \hat\xi_i^*(\breve \kkv_i) \lf[ \hat{\psi}(\kv_1, \ldots, \kv_N) +  \green(\kv_1, \ldots, \kv_N) \: \hat\xi_i(\breve \kkv_i) \ri]	\\
	- \sum_{i < j} \int_{\R^{3N}} \diff \kv_1 \cdots \diff \kv_N \: \chi_R(k_i) \:	\hat\xi_i^*(\breve \kkv_i) \green(\kv_1, \ldots, \kv_N) \: \chi_R(k_j) \: \hat\xi_j(\breve \kkv_j)\,.
}
In the limit $ R \to \infty $ we assume that $\rho^R_i , \, \xi^R_i \rightarrow \xi_i$. Moreover, we extract from the diagonal part of \eqref{ren qform} only the terms not vanishing in that limit
\bmln{
	\sum_{i=1}^N \int_{\R^{3N-3}} \diff \breve \kkv_i \: \lf| \hat\xi_i (\breve \kkv_i) \ri|^2 \lf[ - \frac{(2\pi)^3}{\mu(\al,R)} - \int_{\RT} \diff \kv_i \: \chi_R(k_i) \green(\kv_1, \ldots, \kv_N) \ri] 	\\
	= \sum_{i=1}^N \int_{\R^{3N}}  \diff \breve \kkv_i \: \lf| \hat\xi_i (\breve \kkv_i) \ri|^2 \bigg[ - \frac{(2\pi)^3}{\mu(\al,R)} - 4 \pi R \\	
	+ 2 \pi^2 \bigg[ \frac{m(m+2)}{(m+1)^2} \sum_{j \neq i} k_j^2 + \frac{2m}{(m+1)^2} \sum_{i \neq j} \kv_i \cdot \kv_j + \la \bigg]^{1/2}  + o(1) \bigg].
}
In order to remove the cut-off one is thus forced to set $ \mu \to 0 $ as $ R \to \infty $ and, although several choices are allowed, we set
\beq
	\mu(\al,R) : = - \frac{(2\pi)^3}{4 \pi R + \al},
\eeq
this way canceling the singular term proportional to $- 4\pi R $ contained in the expression above. 

We can now remove the cut-off taking the limit $ R \to \infty $ and so recovering the expression \eqref{form}. Note that we exploit at this stage the fermionic symmetry, which in particular implies that all charges can be expressed in terms of a single function $ \xi $, i.e., 
\beq
	\xi_i(\xv_1,\ldots,\xv_{N-1}) = (-1)^{i+1} \xi(\xv_1,\ldots,\xv_{N-1}),
\eeq 
and $ \xi $ itself is totally antisymmetric under exchange of coordinates. This in turns implies that the sign in front of the off-diagonal term is the opposite than in the bosonic case, implying a completely different behavior of the ground state.


\vs

\end{document}